\documentclass[aps,twocolumn,nofootinbib]{revtex4}

\usepackage{graphicx}
\usepackage{dcolumn}
\usepackage{bm}

\usepackage{amssymb,amsfonts}
\usepackage{epsfig}
\usepackage{epstopdf}
\usepackage[all]{xy}
\usepackage{amsthm}
\usepackage{dcolumn}
\usepackage{hyperref}
\usepackage{url}
\usepackage{dsfont}

\newcommand{\be}{\begin{equation}}
\newcommand{\ee}{\end{equation}}
\newcommand{\bea}{\begin{eqnarray}}
\newcommand{\nn}{\nonumber}
\newcommand{\eea}{\end{eqnarray}}

\def\inbar{\,\vrule height1.5ex width.4pt depth0pt}
\def\IR{\relax{\rm I\kern-.18em R}}
\def\IC{\relax\hbox{$\inbar\kern-.3em{\rm C}$}}

\begin{document}

\title{Gupta-Bleuler quantization for linearized gravity in de Sitter spacetime}

\author{Hamed Pejhan$^{1}$\footnote{pejhan@zjut.edu.cn}}

\author{Mohammad Enayati$^1$}

\author{Jean-Pierre Gazeau$^{2}$\footnote{gazeau@apc.in2p3.fr}}

\author{Anzhong Wang$^{1,3}$\footnote{Anzhong-Wang@baylor.edu}}

\affiliation{$^1$Institute for Theoretical Physics and Cosmology, Zhejiang University of Technology, Hangzhou 310032, China}

\affiliation{$^2$APC, Univ Paris Diderot, Sorbonne Paris Cit$\mbox{\'{e}}$ Paris 75205, France}

\affiliation{$^3$GCAP-CASPER, Physics Department, Baylor University, Waco, TX 76798-7316, USA}

\begin{abstract}
In a recent Letter, we have pointed out that the linearized Einstein gravity in de Sitter (dS) spacetime besides the spacetime symmetries generated by the Killing vectors and the evident gauge symmetry also possesses a hitherto `hidden' local (gauge-like) symmetry which becomes anomalous on the quantum level. This gauge-like anomaly makes the theory inconsistent and must be canceled at all costs. In this companion paper, we first review our argument and discuss it in more detail. We argue that the cancelation of this anomaly makes it impossible to preserve dS symmetry in linearized quantum gravity through the usual canonical quantization in a consistent manner. Then, demanding that all the classical symmetries to survive in the quantized theory, we set up a coordinate-independent formalism \`{a} \emph{la} Gupta-Bleuler which allows for preserving the (manifest) dS covariance in the presence of the gauge and the gauge-like invariance of the theory. On this basis, considering a new representation of the canonical commutation relations, we present a graviton quantum field on dS space, transforming correctly under isometries, gauge transformations, and gauge-like transformations, which acts on a state space containing a vacuum invariant under all of them. Despite the appearance of negative norm states in this quantization scheme, the energy operator is positive in all physical states, and vanishes in the vacuum.
\end{abstract}
\maketitle

\section{Introduction}
In a recent work \cite{Hidden}, with respect to a coordinate-independent approach based on ambient space notations, we have shown that linearized quantum gravity in dS spacetime, constructed through canonical quantization and the usual representation of the canonical commutation relations, suffers from a hitherto `hidden' local (gauge-like) anomaly. More technically, we have shown that the classical theory besides the spacetime symmetries generated by the Killing vectors and the evident gauge symmetry,\footnote{Here, in order to make our discussion explicit, we have used the so-called conformal (global) coordinates,
$$x=(x^0=H^{-1}\tan\rho, (H\cos\rho)^{-1} u), \rho\in]\frac{-\pi}{2},\frac{\pi}{2}[, u\in S^3,$$
in which, the graviton field $h_{\mu\nu}$ can be expressed in terms of the second-rank symmetric tensor spherical harmonics on the three-spheres. [$H$ is the Hubble constant.]}
\begin{eqnarray}\label{gauge}
&h_{\mu\nu} \rightarrow h_{\mu\nu} + 2\nabla_{(\mu}\xi_{\nu)},&
\end{eqnarray}
where $\xi_{\mu}$ is an arbitrary vector field and $\nabla_\mu$ is the covariant derivative, also possesses the additional symmetry
\begin{eqnarray}\label{gauge-like}
&h_{\mu\nu} \rightarrow h_{\mu\nu} + {\cal{E}}_{\mu\nu}\chi,&
\end{eqnarray}
in which ${\cal{E}}_{\mu\nu}$ and $\chi$, respectively, stand for a second-order differential operator (a spin-two projector tensor) and an arbitrary constant function \cite{Hidden}. This hitherto `hidden' gauge-like symmetry, however, becomes anomalous in the quantized theory. Indeed, this additional symmetry of the existing physics, reflected by the subspace generated by the lowest mode (the zero mode) in the set of solutions, reveals that covariant quantization of the dS graviton field inevitably contains a proper quantization of the zero mode of the field. This mode has positive norm, but it violates dS boost invariance as an essential part of dS symmetry. More precisely, under the action of the dS boost generators, it produces all the \emph{negative frequency solutions}\footnote{Recalling the fact that dS spacetime is not stationary and therefore there is \emph{a priori} no natural time coordinate and no natural notion of `positive nor negative frequency' on this spacetime, the term `positive and negative frequency solutions' is used here with respect to the conformal time.} to the field equation.

As is well-known, quite contrary to global anomalies which can be phenomenologically welcomed, generally any local (gauge) symmetry that becomes anomalous makes the theory inconsistent. Therefore, such anomalies must be canceled at all costs (in this regard see for instance \cite{42,13,23,Hooft,Preskill}). This cancelation however imposes strong restrictions on anomalous gauge theories. In our case and with respect to the usual canonical quantization scheme, it seems that the dS symmetry breaking is indeed the price that must be paid for overcoming this difficulty: the natural way out would be to adopt a restrictive version of covariance (for which the action of the dS boost generators is not taken into account) by admitting vacua invariant under a maximal subgroup of the full dS group, the so-called spontaneous symmetry breaking.

In the present paper, we elaborate further on these arguments. However, as dS symmetries are basic symmetries of field dynamics in dS space, our approach to circumvent this local anomaly would be different: we require full covariance as well as causality. Here, we recall that constructing a consistent quantum field theory of gravity in the dS spacetime is of paramount importance, since the symmetry properties of this spacetime can be used as a guideline which greatly helps in the otherwise difficult task of quantizing graviton field in a gravitational background. Indeed, dS spacetime has a privileged status as the unique, maximally symmetric solution to the Einstein equation with positive cosmological constant, which also provides the opportunity of controlling the transition to the flat spacetime by the so-called contraction procedure (see \cite{J. Renaud} and references cited therein). The dS spacetime therefore should at least be respected as an excellent laboratory.

In this sense, it would be pertinent to extend our quantization scheme to a more general context transforming correctly under isometries, gauge transformations, and gauge-like transformations. In this regard, by adopting to this specific situation the content of the previous papers \cite{Hidden,BambaII,BambaI}, we address in this paper the question of constructing a quantum field theory for the linearized Einstein gravity in 3+1-dimensional dS spacetime that be fully covariant according to criteria adapted from the Wightman-G$\mbox{\"{a}}$rding axiomatic for massless fields (Gupta-Bleuler scheme) \cite{Wightman}.

To achieve this goal, the rest of this paper is organized as follows: In Section II, we briefly review the dS machinery. By this we mean a set of definitions and notations concerning geometry and the Linearized Einstein wave equations on one hand, and on the other hand, the corresponding group-theoretical framework. We particularly focus on describing the dS graviton field equation as an eigenvalue equation of the Casimir operators of the dS group, $SO_0(1,4)$. Our formalism, based on ambient space notations, constitutes a coordinate-independent approach to the dS graviton field. It also turns out being a convenient framework to explicitly specify the gauge and the gauge-like degrees of freedom of the theory. Of course, we shall consider the conformal coordinates through this paper to make our construction explicit.

In Section III, we present the smallest, complete, non-degenerate, and dS-invariant space of solutions to the field equation which is called the total space. We prove that this total space is a Krein space. It contains two different types of non-physical modes which are indeed the price to pay for the fully covariance of the theory. The first one appears due to the evident gauge symmetry (\ref{gauge}) and is similar to the non-physical states in gauge quantum field theories in Minkowski space, while the other appears due to the presence of the gauge-like symmetry (\ref{gauge-like}) and is similar to the case of dS minimally coupled scalar field \cite{AllenFolacci,Gazeau1415,de Bievre6230}. [The latter with negative frequency, as already pointed out, is responsible for dS breaking in linearized quantum gravity with respect to the canonical quantization and the usual representation of the canonical commutation relations.] The presence of these non-physical modes naturally leads us to adopt a construction \emph{\`{a} la} Gupta-Bleuler. Actually, for each gauge symmetries of the theory, we have a separate Gupta-Bleuler triplet. The invariant space is defined here according to an indecomposable representation of the dS group carried by these Gupta-Bleuler triplets on the set of solutions. Physical modes, which would be determined up to the gauge and the gauge-like transformations, correspond to the \emph{massless}\footnote{The term ``massless" is used here with respect to conformal invariance and propagation on the dS light-cone.} spin-2 dS UIRs as the central part of the indecomposable representation.

In Section IV, we derive the associated commutator fulfilling the minimal conditions of field equation, i.e. locality and covariance, in closed form: it is expressed in terms of maximally symmetric bitensors (see \cite{AllenJacobson}) in a completely geometric and coordinate-independent form, and found to be finite for points that are not null-related. We indeed point out that the only graviton two-point function in dS space which naturally appears is the commutator that is not of positive type. More precisely, due to the appearance of the anomalous gauge-like symmetry in the usual quantization scheme, any definition \emph{a priori} of a two-point function to construct a graviton field cannot cause a covariant theory.

In Section V, providing a new representation of the canonical commutation relations, the graviton quantum field is given. It is causal and it is covariant in the usual strong sense: $\underline{U}_{\cal G} \underline{h}_{\mu\nu} (X) \underline{U}_{\cal G}^{-1} =  \underline{h}_{\mu\nu} ({\cal G}.X)$, for all ${\cal G}\in SO_0(1,4)$, while $\underline{U}$ stands for the corresponding indecomposable representation of the dS group on the space of states. This implies that the field is defined on the whole dS spacetime. The Fock space carrying this representation is based on the Krein space. In this section, we also discuss that, despite the appearance negative norm states in the quantized theory, no negative energy can be measured: expressions as $\langle \overrightarrow{{\cal{P}}} |T_{00}| \overrightarrow{{\cal{P}}} \rangle$ are always positive in all physical states $|\overrightarrow{{\cal{P}}}\rangle$. This assures a reasonable physical interpretation of the theory.

Finally, further discussion is given in Section VI. We have also supplied some useful identities and mathematical details of calculations in the appendices.

\section{linearized gravity in dS spacetime}

\subsection{Covariant description}
We begin our discussion by considering the Einstein-Hilbert gravity in four spacetime dimensions with positive cosmological constant $\Lambda >0$, while the Lagrangian density is
\begin{eqnarray}\label{action}
{\cal L}_i = \frac{1}{16\pi G}(R-2\Lambda)\sqrt{-g},
\end{eqnarray}
where $G$ is the Newton's constant (we will use units such that $\hbar=c=1$, but $G$ is retained explicitly), $R$ is the Ricci scalar constructed from the metric $g_{\mu\nu}$ and $g$ is the metric determinant. dS space is the unique, maximally symmetric solution to the vacuum Einstein's equations derived from the Lagrangian density (\ref{action}). It is positively curved with fundamental length $H^{-1} = \sqrt{3/\Lambda}$.

To uncover the physics of the theory, one must expand the Lagrangian density (\ref{action}) about the dS background metric that it obviously admits by writing $g_{\mu\nu} = \hat{g}_{\mu\nu} + h_{\mu\nu}$ for dS metric $\hat{g}_{\mu\nu}$ and perturbation $h_{\mu\nu}$. On this basis, the corresponding linearized equations of motion, with respect to the aforementioned conformal coordinates, would be
\begin{eqnarray} \label{Linequ}
&(\Box_H + 2H^2)h_{\mu\nu} - (\Box_H - H^2)\hat{g}_{\mu\nu}h' - 2\nabla_{(\mu}\nabla^\rho h_{\nu)\rho} &\nonumber \\
&+ \hat{g}_{\mu\nu}\nabla^\lambda\nabla^\rho h_{\lambda\rho} + \nabla_\mu\nabla_\nu h'=0,&
\end{eqnarray}
where $\Box_H = \hat{g}_{\mu\nu}\nabla^\mu\nabla^\nu$ is the Laplace-Beltrami operator and $h^\prime = \hat{g}_{\mu\nu}h^{\mu\nu}$. These equations are invariant under the gauge transformation in the form of (\ref{gauge}). Imposing the Lorenz gauge condition upon the metric perturbation, that is, $\nabla^\mu h_{\mu\nu} = e\nabla_\nu h^\prime$, with $e=1/2$, we will partially fix the gauge degree of freedom.

For the sake of argument, we rewrite the field equation (\ref{Linequ}) in a more convenient form defined by the isometric embedding of dS space in a five-dimensional Minkowski space $\mathbb{R}^5$.

\subsection{Embedding space description}
In $\mathbb{R}^5$, with a metric $\eta_{\alpha\beta} = \mbox{diag}(1, -1, -1, -1, -1)$, the 4-dimensional dS manifold can be viewed as a one-sheeted hyperboloid $M_H$ defined by all five-vectors $x^\alpha$ which satisfy: $\eta_{\alpha\beta}x^\alpha x^\beta = -H^{-2}$. The dS metric is given by inducing the natural metric on the hyperboloid
\begin{eqnarray}
ds^2= \eta_{\alpha\beta}dx^{\alpha}dx^{\beta}|_{x^2=-H^{-2}} = {\hat{g}}_{\mu\nu} dX^{\mu}dX^{\nu},
\end{eqnarray}
where $\mu,\nu=0,1,2,3$ and $X^{\mu}$'s refer to the four local spacetime coordinates of $M_H$. This way of describing dS spacetime, constituting the ambient space approach, provides a more suitable framework in which the expressions have a convenient 4+1-Minkowskian form and the symmetries of the theory are easily readable. All this will appear more clearly after writing down the ambient space counterpart of the linearized Einstein equations of motion (\ref{Linequ}).

In the ambient space notations, a tensor field ${\cal K}_{\alpha\beta}$ must satisfy two conditions \cite{Frons1979}, namely:
\begin{itemize}
\item\emph{Homogeneity:} ${\cal K}$ is a homogeneous function of degree $\sigma$ with respect to the ${{\mathbb{R}}}^5$-variables $x^\alpha$, $x_{\alpha}(\partial/\partial x_{\alpha}){\cal K}\equiv x\cdot\partial {\cal K}=\sigma {\cal K}$, with $\sigma\in {\mathbb{R}}$. Note that $\sigma$ is an arbitrarily chosen degree. In the following, we consider $\sigma=0$ for which $\Box_H$ on dS space coincides with the operator $\Box_5\equiv \partial^2$ on $\mathbb{R}^5$ \cite{MVF-GGRT}.
\item\emph{Transversality:} ${\cal K}$ is constrained to be transverse, $x\cdot{\cal K}=0$, which ensures that the direction of ${\cal K}$ lies in the dS space. [In view of the importance of this condition for dS fields, the transverse projection ${T}$ of a tensor field with arbitrary rank is defined as $({T}{\cal K})_{\alpha_1...\alpha_s}=\theta_{\alpha_1}^{\beta_1}...\theta_{\alpha_s}^{\beta_s}{\cal K}_{\beta_1...\beta_s}$, where $\theta_{\alpha\beta}=\eta_{\alpha\beta}+H^2x_\alpha x_\beta$. This projection operator will guarantee the transversality in each index.]
\end{itemize}

In this formalism, the intrinsic tensor field ${\cal K}_{\mu\nu}(X)$ is locally determined by ${\cal K}_{\alpha\beta}(x)$ through the relation
\begin{eqnarray}\label{tranre1}
{\cal K}_{\mu\nu}(X) \Big(\equiv h_{\mu\nu}(X)\Big) = x^\alpha_\mu x^\beta_\nu {\cal K}_{\alpha\beta}(x(X)),
\end{eqnarray}
where $x^\alpha_\mu = \partial x^\alpha/\partial X^\mu$. The covariant derivatives are transformed, for example, as
$$\nabla_\rho\nabla_\lambda h_{\mu\nu} = x_{\rho}^{\alpha}x_{\lambda}^{\beta}x_{\mu}^{\gamma}x_{\nu}^{\sigma}{T}\bar\partial_\alpha {T}\bar\partial_\beta {\cal{K}}_{\gamma\sigma},$$
where $\bar\partial = {T}\partial$ is the transverse derivative in dS space. Following this transformation, $\theta_{\alpha\beta}$ is indeed the only symmetric and transverse tensor which is linked to the dS metric $\hat{g}_{\mu\nu}$.

Considering all of the above identities, the field equation (\ref{Linequ}) takes the form \cite{Frons1979}
\begin{eqnarray}\label{ambicount}
\bar\partial^2{\cal K} - H^2{\cal S}x\partial\cdot{\cal K} - {\cal S}\bar\partial\partial\cdot{\cal K} + \textstyle\frac{1}{2}{\cal S}\bar\partial\bar\partial{\cal K}^\prime \hspace{1.5cm}\nonumber \\
 + \textstyle\frac{1}{2}H^2{\cal S}x\bar\partial{\cal K}^\prime =0,
\end{eqnarray}
where the vector symmetrizer ${\cal S}$ is defined as ${\cal S}(\zeta_\alpha \omega_\beta)=\zeta_\alpha \omega_\beta + \zeta_\beta \omega_\alpha$ and ${\cal K}^\prime$ denotes the trace of ${\cal K}$, and $\partial\cdot {\cal K} = \bar\partial\cdot {\cal K}$ (because, ${\cal K}$ is a transverse tensor).

Here, our aim is to express the field equation (\ref{ambicount}) in terms of the coordinate-independent Casimir operator of the dS group which makes the group theoretical content of the theory explicit. But before that, let us briefly review the definitions and properties of the dS group and its representations that are physically interesting.

\subsection{dS group and representations}
The dS relativity group is the connected Lorentz group $SO_0(1,4)$ of the ambient Minkowski space. The ten infinitesimal generators of the dS group can be split into the orbital and spinorial parts,
\begin{eqnarray}\label{action??}
L^{(r)}_{\alpha\beta}=M_{\alpha\beta}+S^{(r)}_{\alpha\beta},
\end{eqnarray}
where the orbital part is of the form $M_{\alpha\beta} = -i(x_\alpha\partial_\beta - x_\beta\partial_\alpha)$ and $S^{(r)}_{\alpha\beta}$ is the spinorial part acting only on the indices of the rank-$r$ tensors in the following permutational way
$$S_{\alpha\beta}^{(r)}{\cal K}_{\alpha _1 ... \alpha _r} = -i\displaystyle\sum_{i=1}^{r}\Big(\eta_{\alpha\alpha _i}{\cal K}_{\alpha _1...(\alpha _i\rightarrow \beta)...\alpha_r} - (\alpha\leftrightharpoons\beta)\Big).$$
The second-order Casimir operator $Q_r=-\textstyle\frac{1}{2}L^{(r)}_{\alpha\beta}L^{{(r)}\alpha\beta}$ commutes with all generator representatives $L_{\alpha\beta}$ and must therefore, in a given unitary irreducible representation (UIR), be represented by a number, i.e.,
\begin{eqnarray}\label{fieqca}
(Q_r - \langle Q_r\rangle ){\cal K}=0,
\end{eqnarray}
where, with respect to the notation given in \cite{Dix}, we have
\begin{eqnarray}\label{spec}
\langle Q_r\rangle = [-p(p+1)-(q+1)(q-2)],
\end{eqnarray}
in which $p$ and $q$ are two parameters with $2p\in\mathbb{N}$ and $q\in\mathbb{C}$.

According to the possible values for $p$ and $q$, there are three distinct classes of UIRs \cite{Dix, Taka}, i.e. the principal, the complementary and the discrete series. The UIRs associated with our study are those among the discrete series which are classified as two types:
\begin{itemize}
\item The scalar case $\Pi_{p,0}$, with $p=1,2,...$ .
\item The spinorial case $\Pi_{p,q}^\pm$, with $p= 1, 2, ...$ and $q=p, p-1, ..., 1$,
\end{itemize}
in which, the parameter $q$ has a spin meaning. The only physical representations in the sense of Poincar$\acute{\mbox{e}}$ limit are those with $p=q=s$, and the symbol $\pm$ stands for the helicity. These representations are called the massless UIR's.

\subsection{Casimir operators in the field equation}
The action of the second-order Casimir operator on a rank-$2$ tensor field ${\cal K}$ can be written explicitly as follows ($Q_{r=2}\equiv Q_2$)
\begin{eqnarray}\label{acCasi}
Q_2{\cal K} = (Q_0 - 6){\cal K} + 2\eta {\cal K}^\prime + 2{\cal S}x\partial\cdot {\cal K},
\end{eqnarray}
where $Q_0 = -\textstyle\frac{1}{2} M_{\alpha\beta}M^{\alpha\beta} = -H^{-2}\bar\partial ^2$ is the scalar Casimir operator. [The operators $Q_2$, $Q_0$, $L_{\alpha\beta}$ and $\bar\partial$ commute with $x^2$, which means that they are intrinsically defined on the hyperboloid.] With the aid of (\ref{acCasi}), the wave equation (\ref{ambicount}) takes the form \cite{Frons1979}
\begin{eqnarray}\label{Casimcount}
(Q_2 + 6){\cal K} + D_2\partial_2\cdot{\cal K} = 0,
\end{eqnarray}
where $D_2= H^{-2}{\cal S}(\bar\partial - H^2x)$ and $\partial_2\cdot{\cal K}=\partial\cdot{\cal K}-H^2x{\cal K}^\prime-\textstyle\frac{1}{2}\bar\partial{\cal K}^\prime$.

The equation (\ref{Casimcount}) is derivable from the following Lagrangian density
\begin{eqnarray}\label{Lagden}
{\cal L}_i = -\frac{1}{2x^2}{\cal K}\cdot\cdot(Q_2+6){\cal K}+\frac{1}{2}(\partial_2\cdot{\cal K})^2,
\end{eqnarray}
where `$\cdot\cdot$' denotes total contraction. The Lagrangian density (\ref{Lagden}) is invariant under the ambient space counterpart of (\ref{gauge}), that is, ${\cal K}\rightarrow {\cal K} + D_2\lambda$, where $\lambda$ is an arbitrary vector field. We now need to break this gauge symmetry. To do this, we use the gauge-fixing term ${\cal L}_{gf}=(1/2a)(\partial_2\cdot{\cal K})^2$, where `$a$' is an arbitrary constant: note that the ambient space counterpart of the Lorenz gauge condition is $\partial_2\cdot{\cal K} = (b-\textstyle\frac{1}{2})\bar\partial{\cal K}^\prime$, with $b=\textstyle\frac{1}{2}$. Finally, a variation of the action integral $\int({\cal L}_i + {\cal L}_{gf})d\sigma$, where $d\sigma$ is the volume element in dS space, leads to the following equation
\begin{eqnarray}\label{cCasimcount}
(Q_2 + 6){\cal K} + cD_2\partial_2\cdot{\cal K} = 0,
\end{eqnarray}
where $c = (1+a)/a$ is a gauge-fixing parameter.

It is obvious that for $c=1$, the field equation (\ref{cCasimcount}) is fully gauge invariant. In the following, we will explicitly see that the simplest (or optimal) choice of $c$ is precisely $c=2/5\equiv c_\ell$.\footnote{Generally, for a spin-$s$ field, the simplest choice is $c_\ell=2/(2s+1)$ \cite{Gazeau507,Gazeau1985g}.} However, gauge freedom still exists by any choice of $c\neq 1$: the choice of the vector fields $\lambda$ is only limited to what can be achieved by\footnote{Note that, $\partial_2\cdot D_2 \lambda = -(Q_1+6)\lambda$ and $Q_2D_2\lambda=D_2Q_1\lambda$.}
\begin{eqnarray}\label{gafieq}
(1-c)D_2(Q_1+6)\lambda = 0.
\end{eqnarray}
This means that the Lorenz gauge partially fixes the gauge degree of freedom. In this sense, the physical solutions which are indeed among the divergencelessness solutions (i.e., the solutions to $(Q_2+6){\cal K} = 0$) constitute a part of an indecomposable structure while the gauge solutions are determined by Eq. (\ref{gafieq}) and the non-zero divergence solutions, called the scalar solutions, obey
\begin{eqnarray}\label{diivequ}
(1-c)(Q_1+6)\partial_2\cdot{\cal K} = 0.
\end{eqnarray}

In this indecomposable structure the physical solutions carry the massless spin-2 representations $\Pi_{2,2}^\pm$, while the gauge and the scalar sectors of solutions carry the dS finite-dimensional representation labeled by the pair $(p=2,q=-1)$ which is Weyl equivalent to $\Pi_{2,2}^\pm$.\footnote{A necessary condition that the representations $\Pi_{2,2}^\pm$ combine with other ones to form indecomposable representations is that the latter are Weyl equivalent. [If two representations are Weyl equivalent, then they share same Casimir eigenvalue.]} As a matter of fact, regarding the two possible degrees of homogeneity $\sigma=0$ and $\sigma=-3$, which will replace in dS space the two helicities of the graviton field, the gauge and the scalar sectors carry the dS finite-dimensional representation $(p=2,q=-1)$ and the non-unitary infinite-dimensional representation $(p=1,q=3)$, respectively (see the details in Refs. \cite{Fronsdal-1987,Gazeau-1988-2533}). In this study, however, for the sake of simplicity we only consider $\sigma=0$, i.e. $x\cdot\partial {\cal K}=0$, and consequently the representation $(p=2,q=-1)$: here, by abuse of notation, we denote this representation by $\Pi_{2,-1}$. Of course, our result can be simply generalized to the case $\sigma=-3$.

\section{Space of solutions}
In this section, we define the total space of solutions, that is, the complete, non-degenerate and fully invariant space of solutions to the field equation (\ref{cCasimcount}). Then, on this total space, we present the Gupta-Bleuler triplets carrying the indecomposable structure for the dS group unitary representation appearing in the linearized gravity in dS spacetime.

\subsection{Total space}
We begin by producing a recurrence formula expressing the tensor field ${\cal K}$ of rank-$2$ in terms of the tensors of lower ranks to set up general solutions verifying (\ref{cCasimcount}). Such recurrence formula involves operators that obey commutation/intertwining rules with $L_{\alpha\beta}^{(2)}$ and the Casimir operator $Q_2$.

In such a construction, the contraction of the transverse projector $\theta$ with a constant polarization five-vector $Z$, i.e. ${\cal S} \theta\cdot Z$, is a key piece in the sense that permits one to define an operator that makes a symmetric transverse tensor field ${\cal K}$ of rank-$2$ from a transverse tensor field $K$ of rank-$1$ \cite{Gazeau-1988-2533}. The commutation rule between ${\cal S} \theta\cdot Z K$ and $Q_2$ is given by
\begin{eqnarray}
Q_2{\cal S}\theta\cdot ZK={\cal S}\theta\cdot Z(Q_1-4)K \hspace{1.5cm}\nn\\
 -2H^2D_2x\cdot ZK +4\theta Z\cdot K.\nonumber
\end{eqnarray}
Regarding the above identity, now we need to calculate the commutation relations between $Q_2$ and the new elements $\theta\phi$ and $D_2 \widetilde{K}$, in which the operators $\theta$ and $D_2$ make a symmetric transverse tensor field ${\cal K}$ of rank-$2$ from a tensor field $\phi$ of rank-$0$ and a transverse tensor field $\widetilde{K}$ of rank-$1$, respectively. For these new elements, we have
\begin{eqnarray}
Q_2\theta\phi=\theta Q_0\phi, \nonumber
\end{eqnarray}
\begin{eqnarray}
Q_2D_2\widetilde{K} = D_2 Q_1 \widetilde{K}.\nonumber
\end{eqnarray}
The above identities explicitly reveal that the elements of the forms ${\cal S}\theta\cdot Z K$, $\theta\phi$, and $D_2\widetilde{K}$ constitute a closed set under the action of $Q_2$. In this sense, we can obtain the tensor field ${\cal K}$ in a dS-invariant way in terms of $K$, $\phi$ and $\widetilde{K}$,
\begin{eqnarray}\label{recuform}
{\cal K} = {\cal S}\theta\cdot Z K + \theta\phi + D_2\widetilde{K}.
\end{eqnarray}

Substituting the above recurrence formula into the field equation (\ref{cCasimcount}) and writing the tensors $\phi$ and $\widetilde{K}$ in a completely dS-invariant manner in terms of $K$ (see the details in \cite{BambaII}), the tensor field ${\cal K}$ reads
\begin{eqnarray}\label{csolu}
{\cal K} = {\cal K}^{c=c_\ell} + \frac{c_\ell - c}{1-c}D_2(Q_1+6)^{-1}\partial_2\cdot{\cal K}^{c=c_\ell},\; c\neq 1
\end{eqnarray}
in which
\begin{eqnarray}\label{clsolu}
{\cal K}^{c=c_\ell} &=& \Big(-\frac{2}{3}\theta Z\cdot K +{\cal S}\bar{Z}K + \frac{1}{27} H^2 D_2D_1 Z\cdot K \nonumber\\
&& + \frac{1}{3 }H^2 D_2 x\cdot Z K \Big) + \frac{1-c_\ell}{9} D_2\partial_2\cdot {\cal K}^{c=c_\ell},\nonumber\\
& = & [\mbox{$c$-independent part}] + \frac{1-c_\ell}{9} D_2\partial_2\cdot {\cal K}^{c=c_\ell},\;\;\;\;\;\;\;
\end{eqnarray}
where $D_1 = H^{-2}\bar\partial$, $\bar{Z}= {T}Z = \theta\cdot Z $, and
\begin{eqnarray}\label{K VECTOR}
(Q_1 + 2)K = 0, \;\;\; \partial\cdot K =0.
\end{eqnarray}

Note that, (\emph{i}) For $c\neq 1$, the trace of ${\cal K}$ is zero, ${\cal K}^\prime = 0$, then we have $\partial_2\cdot{\cal K} = \partial\cdot{\cal K}$ \cite{GaHaMu}. (\emph{ii}) The gauge solutions $D_2\lambda$ only appear coupled to the scalar part, $D_2\partial_2\cdot {\cal K}^{c=c_\ell}$. (\emph{iii}) The last term on the right-hand side of (\ref{csolu}) is responsible for the appearance of logarithmic divergences in the field solutions which implies reverberation inside the light cone. This term can be eliminated by adjusting $c$ to $c_\ell$ (the simplest structure). With this choice of the gauge-fixing parameter even the physically irrelevant gauge modes propagate only on the dS light cone \cite{Flato415,Gazeau523}: indeed, the subscript `$\ell$' in $c_\ell$ stands for this fact.

Interestingly, one can pursue the above procedure and write the tensor field $K$ of rank-$1$ in a dS-invariant manner in terms of a tensor field $\Phi$ of rank-$0$ to set up general solutions verifying (\ref{K VECTOR}) \cite{BambaII},
\begin{eqnarray}\label{vectorfield}
K = \Big( \bar{Z^\prime} - \textstyle\frac{1}{2}D_1(Z^\prime\cdot\bar\partial + 2H^2x\cdot Z^\prime)\Big) \Phi(x),
\end{eqnarray}
where $Z^\prime_\alpha$ is another constant five-vectors and
\begin{eqnarray}\label{Q_0 phi}
Q_0 \Phi(x) = -H^{-2}\Box_H \Phi(x) = 0.
\end{eqnarray}

In this sense, the general solutions to the field equation (\ref{cCasimcount}) can remarkably be written as the resulting action of a dS-invariant, spin-two projector ${\cal E}_{\alpha\beta}(x,\partial,Z,Z^\prime)$ on a massless minimally coupled scalar field $\Phi(x)$ (the structure function),
\begin{eqnarray}\label{general solu}
{\cal K}_{\alpha\beta} = {\cal E}_{\alpha\beta}(x,\partial,Z,Z^\prime) \Phi(x), \;\;\; c\neq 1.
\end{eqnarray}

Regarding the structure function obeying (\ref{Q_0 phi}), there exists a continuous family of simple solutions, the so-called coordinate-independent dS plane waves, as \cite{BrosPRL,BrosMoschella}
\begin{eqnarray}\label{planew}
\Phi (x) = (Hx\cdot \xi)^{-3},
\end{eqnarray}
where the vectors $\xi\in\mathbb{R}^5$ lying in the null-cone ${\cal C}=\{\xi\in\mathbb{R}^5;\;\xi^2=0\}$.\footnote{Substituting (\ref{planew}) into (\ref{general solu}) and after a direct calculations, along the lines given in Ref. \cite{MVF-GGRT}, one can present the general solutions (\ref{general solu}) in the form of ($c\neq 1$)
$${\cal K}_{\alpha\beta}(x,\xi,Z,Z^\prime) = {\cal E}_{\alpha\beta}(x,\xi,Z,Z^\prime) (Hx\cdot \xi)^{-3}.$$
This presentation makes the degree of homogeneity of the general solutions apparent: the waves ${\cal K}_{\alpha\beta}(x,\xi,Z,Z^\prime)$, as functions on $\mathbb{R}^5$, are homogeneous with degree zero (note that, $H(x)=1/\sqrt{-x\cdot x}$).} In the conformal coordinates introduced above, these solutions (\ref{planew}) take the form (see Appendix (\ref{dS wave planes}))
\begin{eqnarray}\label{phiint}
\Phi_{Llm}(\rho, u) = i^{L+3} e^{-iL\rho}  \frac{\Gamma(L+3)}{(L+1)! \Gamma(3)}\hspace{2cm} \nonumber\\
\times {_2F_1}(-1,L;L+2;-e^{-2i\rho})Y_{Llm}(u),
\end{eqnarray}
where $Y_{Llm}(u)$ are the hyperspherical harmonics on $S^3$, with $(L, l, m)\in \mathbb{N}\times\mathbb{N}\times\mathbb{Z}$, $0\leq l\leq L$ and $-l\leq m\leq l$.

With the aid of (\ref{tranre1}), the solutions to the equation of motion in the conformal coordinates can be expressed as
\begin{eqnarray}\label{amitoin}
{\cal K}_{\mu\nu}^{(L,l,m)} = x_\mu^\alpha x_\nu^\beta {\cal E}_{\alpha\beta}\Phi_{Llm}(\rho, u) \equiv {\cal E}_{\mu\nu}\Phi_{Llm}(\rho, u).
\end{eqnarray}

We define $V^{c}$, the total space of solutions, as the complete set of solutions in the form (\ref{amitoin}) which are square integrable with respect to the following dS-invariant bilinear form (or inner product) \cite{GaHaMu}
\begin{eqnarray} \label{B1}
&\langle {\cal K}_1,{\cal K}_2\rangle = \frac{i}{H^2} \int_{S^3,\rho=0} [({{\cal K}_1})^\ast\cdot\cdot\partial_\rho{{\cal K}_2} &\nonumber\\
&- 2 c ((\partial_\rho x)\cdot{({{\cal K}_1})}^\ast)\cdot(\partial\cdot{{\cal K}_2}) - (1^\ast \leftrightharpoons 2)]d\sigma(u),&
\end{eqnarray}
in which ${\cal K}_1$ and ${\cal K}_2$ are two arbitrary modes. A closer look at the behavior of the above inner product however reveals that $V^{c}$ suffers from a degeneracy associated with the lowest mode (the mode corresponding to $L=0$ in (\ref{amitoin})). In fact, considering the behavior of the hypergeometric functions \cite{Magnus},
\begin{eqnarray}
_2F_1(-1,L;L+2;-e^{-2i\rho})=1-\frac{L}{L+2}e^{-2i\rho},\nonumber
\end{eqnarray}
for which the $L=0$ mode is determined by ${\cal E}_{\mu\nu}\Phi^{}_{0,0,0}$, where $\Phi^{}_{0,0,0}$ is a constant function, one can easily see that this mode is orthogonal to the entire set of solutions including itself. Of course, this is not an accidental degeneracy (due to a mathematical artifact). Indeed, with respect to our formalism it is obvious that, whatever the value of gauge-fixing parameter $c$ is, the theory (the inner product and equation of motion) is invariant under the gauge-like transformations (\ref{gauge-like}). This additional symmetry of the dS linearized Einstein gravity directly implies that the null-norm subspace, generated by $L=0$, should be viewed as a space of `gauge' states.

In this sense, we need to delve more deeply into the lowest graviton mode determined by $L=0$ in the set of solutions in order to circumvent the associated degeneracy problem. Accordingly, by solving Eq. (\ref{Q_0 phi}) directly for $L=0$ (see \cite{Gazeau1415}), we obtain the following two independent solutions
\begin{eqnarray}\label{twoindsol}
{\cal K}_{g_1} = {\cal E}_{\mu\nu}\Big(\frac{H}{2\pi}\Big),\;{\cal K}_s = {\cal E}_{\mu\nu}\Big(-\frac{iH}{2\pi}[\rho+(1/2)\sin 2\rho]\Big),\nonumber
\end{eqnarray}
where the constants are chosen to have $\langle {\cal K}_{g_1}, {\cal K}_s\rangle = 1$. Note that, (\emph{i}) Both ${\cal K}_{g_1}$ and ${\cal K}_s$ are null norm solutions. (\emph{ii}) ${\cal K}_{g_1}$ is a solution to (\ref{cCasimcount}) which has been already appeared in the gauge-like symmetry (\ref{gauge-like}).

On this basis, we define the `true' normalized zero mode, the modified one, as
\begin{eqnarray}\label{modifzmo}
{\cal K}_{\mu\nu}^{(0,0,0)} = {\cal K}_{g_1} + \frac{1}{2}{\cal K}_s,\;\;\; \langle {\cal K}^{(0,0,0)}, {\cal K}^{(0,0,0)}\rangle = 1.
\end{eqnarray}
Including this mode, we have a non-degenerate set of normalized solutions ${\cal K}_{\mu\nu}^{(L,l,m)}$, with $L\geq0$, but the space generated by these modes is not closed under the action of the dS group. More precisely, applying the dS boost generators on ${\cal K}^{(0,0,0)}$ produces the whole positive and negative frequency solutions to the equation of motion. [Obviously, one cannot drop out the $L=0$ mode from the set of solutions, because it leaves us with a non-complete set of solutions violating the gauge-like symmetry (\ref{gauge-like}) of the existing physics.] For instance, under the action of the dS boost generator $(L_{03} + iL_{04})$, we have
\begin{widetext}
\begin{eqnarray}\label{base}
(L_{03} + iL_{04}){\cal K}_{\mu\nu}^{(0,0,0)} &=& \underbrace{\Big((L_{03} + iL_{04}){\cal E}_{\mu\nu}\Big)\Big(\frac{H}{2\pi} -\frac{iH}{4\pi}[\rho+\frac{1}{2}\sin 2\rho] \Big)}_{\mbox{`invariant terms'}}\nonumber\\
&& + {\cal E}_{\mu\nu}\Big( \underbrace{(M_{03} + iM_{04})(\frac{H}{2\pi})}_{=0} + \underbrace{(M_{03} + iM_{04})(-\frac{iH}{4\pi}[\rho+\frac{1}{2}\sin 2\rho])}_{=(\frac{-\sqrt{6}}{4}) (i\Phi_{1,0,0} +i \Phi^*_{1,0,0} +\Phi_{1,1,0} +\Phi^*_{1,0,0})} \Big)\nonumber\\
&=& \mbox{`invariant terms'} - \frac{\sqrt{6}}{4} \Big({\cal K}_{\mu\nu}^{(1,1,0)} + i{\cal K}_{\mu\nu}^{(1,0,0)} + (1+i)({\cal K}_{\mu\nu}^{(1,0,0)})^\ast \Big).
\end{eqnarray}
\end{widetext}
Note that, (\emph{i}) $({\cal K}_{\mu\nu}^{(1,0,0)})^\ast$ refers to the complex conjugate of ${\cal K}_{\mu\nu}^{(1,0,0)}$. (\emph{ii}) According the procedure that has been considered to produce the spin-two projector ${\cal E}_{\mu\nu}$, it is trivial that the elements of ${\cal E}_{\mu\nu}$ for a given $c$ remain in ${\cal E}_{\mu\nu}$ under the group action. (\emph{iii}) To see the behavior of the scalar structure function $\Phi$ under the action of the dS boost generators, one can refer to Ref. \cite{Gazeau1415}.

With respect to the above arguments, it is now clear that the total space of solutions $V^{c}$, if one needs the full invariance of the theory, must be extended to include all the positive and negative frequency solutions to the field equations, i.e.
\begin{eqnarray}\label{compl}
V^c = {\cal V}^{+}\oplus {\cal V}^{-}\oplus {\cal N}\oplus {\cal S},
\end{eqnarray}
in which
\begin{eqnarray}\label{decomp}
{\cal N} &=& \Big\{ c_g{\cal E}_{\mu\nu}\Big(\frac{H}{2\pi}\Big) \Big\}, \nn\\
{\cal S} &=& \Big\{ c_s{\cal E}_{\mu\nu}\Big(-i\frac{H}{2\pi}[\rho + \frac{1}{2}\sin 2\rho ]\Big) \Big\}, \nn\\
{\cal V}^{+}&=& \Big\{ \sum_{Llm,L\geq1} c_{Llm}({\cal E}_{\mu\nu}\Phi_{Llm}); \sum_{Llm,L\geq1}| c_{Llm} |^2<\infty \Big\},\nn\\
{\cal V}^{-} &=& \Big({\cal V}^{+}\Big)^\ast,
\end{eqnarray}
with $c_g, c_s, c_{Llm}\in \mathds{C}$. The following relations demonstrate the behavior of each part of the space of solutions under the action of the dS group
\begin{eqnarray}\label{transf}
U_{\cal G}&:& {\cal N}\longrightarrow {\cal N},\nn\\
U_{\cal G}&:&{\cal S}\longrightarrow {\cal S}\oplus {\cal V}^{+}\oplus {\cal V}^{-}\oplus {\cal N},\nn\\
U_{\cal G}&:&{\cal V}^{+}\longrightarrow {\cal V}^{+}\oplus {\cal N},\nn\\
U_{\cal G}&:&{\cal V}^{-}\longrightarrow {\cal V}^{-}\oplus {\cal N},
\end{eqnarray}
where, for an arbitrary element ${\cal G}$ of the dS group, $U_{\cal G}$ is the dS natural representation on the space of solutions. To see the point lying behind the above transformations, beside Eq. (\ref{base}), for instance one should also consider
\begin{eqnarray}
(L_{03} + iL_{04}){{\cal K}}_{\mu\nu}^{(1,0,0)}= \mbox{`invariant terms'}\hspace{2.6cm} \nonumber\\
+ {{\cal E}}_{\mu\nu}\Big((M_{03} + iM_{04})\Phi_{1,0,0}\Big)\nonumber\\
= ... -i \frac{4}{\sqrt{6}} {\cal K}_{\mu\nu}^{2,1,0} + {\cal K}_{\mu\nu}^{2,0,0} + \frac{3}{2\sqrt{6}}{\cal{K}}_{g_1}.\nonumber
\end{eqnarray}

\subsection{Gupta-Bleuler triplets}
The above arguments reveals that the quantum field theory formulated through canonical quantization and the usual representation of the canonical commutation relations suffers from an anomalous symmetry associated with the (local) gauge-like symmetry (\ref{gauge-like}), for which one has to deal with the propagation of negative frequency states ${\cal K}_{\mu\nu}^\ast$ in the quantized theory. The presence of this local anomaly is not a gauge artifact, no matter how we choose the gauge-fixing parameter $c$, and in general makes it impossible to implement the Gauss constraint on the physical states \cite{42,13,23}. For this reason, one of the most requirements for a consistent dS linearized quantum gravity is that this gauge-like symmetry of the system must be free of anomaly. As already pointed out, with respect to the usual canonical quantization, this requirement can be successfully fulfilled by weakening the covariance condition based on a maximal subgroup of the full dS group, for example $SO(4)$, for which the dS boost invariance is not taken into account. In this paper, however, our approach is different. We need all the classical symmetries to survive in the quantized theory. To achieve this goal, we consider a rather straightforward application of the Gupta-Bleuler formalism, known to be well adapted to treat models with gauge symmetries, to avoid the symmetry breaking altogether.

On this basis, we categorize the total space $V^{c}$ (see (\ref{compl})) as a chain of the solutions, known as the Gupta-Bleuler triplet, in the following form
\begin{eqnarray}\label{trip2}
V_{g_1}\subset V_1 \subset V^{c},
\end{eqnarray}
in which the space $V_{g_1}\equiv{\cal N}$ is the space of gauge-like solutions. On the other hand, we have $V_1 \equiv{\cal V}^{+}\oplus{\cal N}$ which is a space of positive frequency solutions to the field equation equipped with the degenerate inner product: $\langle{\cal K}_{g_1}, {\cal K}_{g_1}\rangle=0$ and $\langle{\cal K}_{g_1}, {\cal K}\rangle=0$, for all ${\cal K}\in {V_1}$. In this construction, the coset space $V^c/V_1$ contains auxiliary solutions which are of negative frequency. These non-physical solutions indeed allow one to overcome in a totally covariant way the aforementioned zero-mode problem.

Here we must underline that the appearance of the above structure (\ref{trip2}) is not the whole story: besides the gauge-like degrees of freedom there are also the degrees of freedom due to the evident gauge invariance. Indeed, a closer look at the subspaces $V_{g_1}$, $V_1/V_{g_1}\equiv {\cal V}^{+}$ and $V^c/V_1\equiv {\cal V}^{-}\oplus {\cal S}$ reveals that a further Gupta-Bleuler triplet appears inside each of them. Let us start with the quotient space of positive frequency solutions to (\ref{cCasimcount}), i.e. $V_1/V_{g_1}\equiv {\cal V}^{+}$. Considering (\ref{gafieq}), (\ref{diivequ}) and the associated discussions in section II, one can realize three main types of solutions in ${\cal V}^{+}$: the divergencelessness type, the evident gauge type, and the solutions which are not divergenceless. On this basis, in the quotient space ${\cal V}^{+}$, one can single out the invariant, but not invariantly complemented, subspace ${V^+_2}\subset {\cal V}^{+}$ made up of elements which are divergenceless, and therefore, are $c$ independent. The non-zero divergence solutions and the evident gauge fields ${\cal K}^+_{g_2}$, respectively, belong to the subspaces ${\cal V}^{+}/V^+_2$ and $V^+_{g_2}$. The latter is an invariant, but again not invariantly complemented, subspace $V^+_{g_2}\subset{V^+_2}$. Note that, the gauge solutions ${\cal K}^+_{g_2}$ are orthogonal to every element in ${V^+_2}$ including themselves \cite{GaHaMu}: $\langle{\cal K}^+_{g_2}, {\cal K}^+_{g_2}\rangle=0$ and $\langle{\cal K}^+_{g_2}, {\cal K}^+\rangle=0$, for all ${\cal K}^+\in {V^+_2}$. [The symbol `$+$' stands for the positivity of frequency in the corresponding subspace of solutions.]

This further chain of solutions inside $V_1/V_{g_1}\equiv {\cal V}^{+}$ can be demonstrated as
\begin{eqnarray}\label{trip1}
V^+_{g_2}\subset {V^+_2}\subset {\cal V}^{+}.
\end{eqnarray}
This triplet carries the following indecomposable structure for the unitary representation of the dS group\footnote{See Eqs. (\ref{gafieq}) and (\ref{diivequ}) and the associated discussions in section II.}
\begin{eqnarray}\label{indestru}
\underbrace{{\Pi_{2,-1}}}_{{\cal V}^{+}/V^+_2}\;\;\rightarrow\;\; \underbrace{\Pi_{2,2}^{+}
\oplus\Pi_{2,2}^{-}}_{V^+_2/V^+_{g_2}}\;\;\rightarrow\;\; \underbrace{{\Pi_{2,-1}}}_{V^+_{g_2}}
\end{eqnarray}
The arrows show the leaks under the dS group action.

Alternately, we could arrive at a comparable result, i.e., the similarity in appearance of the above structure (see (\ref{trip1}) and (\ref{indestru})), for each subspaces $V^c/V_1\equiv {\cal V}^{-}\oplus {\cal S}$ and $V_{g_1}\equiv{\cal N}$ separately. Indeed, $V^c/V_1$ carries the same indecomposable representation as (\ref{indestru}). Of course, as already mentioned, the subspace $V^c/V_1$ contains auxiliary solutions which are of negative frequency. On the other hand, the space $V_{g_1}\equiv{\cal N}$ carries the trivial representations of the dS group associated with (\ref{indestru}).\footnote{Under the action of the dS group (\ref{action??}), we have $$L\Big({\cal{E}}(\frac{H}{2\pi})\Big) = \Big(L{\cal{E}}\Big)(\frac{H}{2\pi}) + {\cal{E}} \Big(\underbrace{M(\frac{H}{2\pi})}_{=0}\Big) = \Big(L{\cal{E}}\Big)(\frac{H}{2\pi}).$$}

Taking all of the above into consideration, the complete indecomposable group representation structure appearing in the dS linearized gravity which is carried by these sets of Gupta-Bleuler triplets altogether (see (\ref{trip1}) and (\ref{trip2})) reads
\begin{eqnarray}\label{complete}
\begin{array}{ccccccc}
{V_{g_1}:} & \;{{\widetilde{\Upsilon}_0}} & \rightarrow & {{\;\;\;\;\;\;\Upsilon_0\;\;\;\;\;\;}} & \rightarrow  &{{\widetilde{\Upsilon}_0}}\; & \\
& \uparrow &  &\uparrow &  &\uparrow &\\
{V_1/V_{g_1}:} &\; {\Pi_{2,-1}} &  \rightarrow & ``{{\Pi_{2,2}^{+} \oplus\Pi_{2,2}^{-}}}" & \rightarrow & {\Pi_{2,-1}} \;&  \\
&  \uparrow & &\uparrow & &\uparrow &\\
{V^c/V_1:} &\; \underbrace{{{\Pi_{2,-1}}}}_{V^c/V_2} &  \rightarrow & \underbrace{{{\Pi_{2,2}^{+} \oplus\Pi_{2,2}^{-}}}}_{V_2/V_{g_2}} & \rightarrow & \underbrace{{{\Pi_{2,-1}}}}_{V_{g_2}}\; &
\end{array}
\end{eqnarray}
The vertical arrows demonstrate the leaks under the dS group action as well, and they can be understood by the transformations (\ref{transf}), while $\Upsilon_0$ and $\widetilde{\Upsilon}_0$, respectively, refer to the trivial UIRs of the dS group and the corresponding Weyl equivalent finite-dimensional representations associated with the eigenvalue of the Casimir operator $\langle Q_r\rangle= -6$. Note that, the space of physical graviton modes corresponds to the dS UIRs denoted between quotations symbol in the above indecomposable representations. The physical space is indeed the invariant subspace in the quotient space $V_2/V_{g_2}$ (the $c$-independent part) which contains no negative norm modes, more precisely, $\Big(V_2/V_{g_2}\Big) \bigcap \Big(V_1/V_{g_1}\Big)$.

At the end, to summarize our discussion in this section, it would be convenient to express the total space of solutions, which is indeed a Krein space, in a more applicable form as
\begin{eqnarray}
V^c =  {\cal H}\oplus {\cal H}^\ast,
\end{eqnarray}
while
\begin{eqnarray}
{\cal H} = \Big\{ \sum_{Llm,L\geq0} c_{Llm}({\cal E}_{\mu\nu}\Phi_{Llm}); \sum_{Llm,L\geq0}| c_{Llm} |^2<\infty \Big\}.\nonumber
\end{eqnarray}
In this way, the invariant subspace of physical states reads
\begin{eqnarray}\label{phys-space}
V_{phys} \equiv \Big(V_2/V_{g_2}\Big) \bigcap \Big(V_1/V_{g_1}\Big) = \hspace{2.5cm}\nonumber\\
\Big\{ \sum_{Llm,L\geq 1} c_{Llm}({\cal E}^{\frac{2}{{g_2}}}_{\mu\nu}\Phi_{Llm}); \sum_{Llm,L\geq 1}| c_{Llm} |^2<\infty \Big\},
\end{eqnarray}
in which ${\cal E}^{\frac{2}{{g_2}}}_{\mu\nu}$ stands for the projection tensor associated with the invariant subspace $V_2/V_{g_2}$, denoted by the $c$-independent part in (\ref{clsolu}). Again, the zero mode ($L=0$) does not contribute to the physical space of solutions because if it was included, the set of physical modes would be transformed into modes of negative frequency violating unitarity. In particular this means that only the $L\geq1$ modes will contribute to any physical quantity.

\section{Graviton two-point function}
In this section, with respect to the discussions given by Allen and Jacobson in their seminal paper \cite{AllenJacobson}, we aim to find the graviton two-point function in closed form in terms of maximally symmetric bitensors. The bitensors are functions of two points $x$ and $x^\prime$, and behave like tensors under coordinate transformations at each point. If they respect the dS invariance, we call them maximally symmetric.

As proved in Appendix (\ref{bitensors}), any maximally symmetric rank-2 bitensor can be presented in the ambient space formalism in terms of the three basic bitensors $\theta \theta^{\prime}$, $\theta\cdot\theta^{\prime}$ and $D_2D^{\prime}_2$. Therefore, the graviton two-point function can be written as
\begin{eqnarray}\label{2point 00}
{\cal{W}}_{\alpha\beta\alpha^{\prime}\beta^{\prime}}(x,x^{\prime})=
\theta_{\alpha\beta}\theta^{\prime}_{\alpha^{\prime}\beta^{\prime}}{\cal{W}}_0(x,x^{\prime})\hspace{2.6cm}\nn\\
+{\cal{S}}{\cal{S}}^{\prime}\theta_\alpha\cdot\theta^{\prime}_{\alpha^{\prime}}{\cal{W}}_{1\beta\beta^{\prime}}(x,x^{\prime})
+{D_2}_{\alpha}{D^{\prime}_2}_{\alpha^{\prime}}{\cal{W}}_{g\beta\beta^{\prime}}(x,x^{\prime}),\;\;\;
\end{eqnarray}
where ${\cal{W}}_0$ is a biscalar, ${\cal{W}}_1$ and ${\cal{W}}_g$ are two bivectors. Note that the primed operators act only on the primed coordinates and vise versa, so that $D_2D^{\prime}_2= D^{\prime}_2D_2$.

The above two-point function has to satisfy the following requirements:
\begin{itemize}
\item{\emph{Indefinite sesquilinear form.} For any test function $f_{\alpha\beta}$ in the space of functions $C^{\infty}$ with compact support in $M_H$, an indefinite sesquilinear form would be \cite{BrosMoschella}
\begin{eqnarray}
\int_{M_H\times M_H}f^{\ast\alpha\beta}(x){\cal{W}}_{\alpha\beta\alpha^\prime\beta^\prime}(x,x^\prime)f^{\alpha^\prime\beta^\prime}(x^\prime)d\sigma (x)d\sigma (x^\prime).\nonumber
\end{eqnarray}}
\item{\emph{Covariance.}
\begin{eqnarray}
({\cal G}^{-1})_\alpha^\gamma({\cal G}^{-1})_\beta^\delta{\cal{W}}_{\gamma \delta \gamma\prime \delta\prime}({\cal G}x,{\cal G}x^\prime){\cal G}_{\alpha^\prime}^{\gamma\prime}{\cal G}_{\beta^\prime}^{\delta\prime}={\cal{W}}_{\alpha\beta\alpha^\prime\beta^\prime}(x,x^\prime),\nonumber
\end{eqnarray}
for all ${\cal G}\in SO_0(1,4)$.}
\item{\emph{Locality.} For every space-like separated pair $(x,x^\prime)$, i.e. $x\cdot x^\prime>-H^{-2}$,
\begin{eqnarray}
{\cal{W}}_{\alpha\beta\alpha^\prime\beta^\prime}(x,x^\prime)={\cal{W}}_{\alpha^\prime\beta^\prime\alpha\beta}(x^\prime,x).\nonumber
\end{eqnarray}}
\item{\emph{Index symmetrizer.}
\begin{eqnarray}
{\cal{W}}_{\alpha\beta\alpha^{\prime}\beta^{\prime}}(x,x^{\prime})= {\cal{W}}_{\alpha\beta\beta^{\prime}\alpha^{\prime}}(x,x^{\prime}) = {\cal{W}}_{\beta\alpha\alpha^{\prime}\beta^{\prime}}(x,x^{\prime}).\nonumber
\end{eqnarray}}
\item{\emph{Transversality.}
\begin{eqnarray}
x^\alpha{\cal{W}}_{\alpha\beta\alpha^{\prime}\beta^{\prime}}(x,x^{\prime})= 0 = {x^{\prime}}^{\alpha^{\prime}}{\cal{W}}_{\alpha\beta\alpha^{\prime}\beta^{\prime}}(x,x^{\prime}).\nonumber
\end{eqnarray}
This automatically results in the transversality condition on ${\cal W}_1$ and ${\cal W}_g$.}
\item{\emph{Tracelessness.}
\begin{eqnarray}
{({\cal{W}})}^\alpha_{\alpha\alpha^{\prime}\beta^{\prime}}(x,x^{\prime})= 0 = {{({\cal{W}})}_{\alpha\beta\alpha^{\prime}}}^{\alpha^{\prime}}(x,x^{\prime}).\nonumber
\end{eqnarray}
This directly yields
\begin{eqnarray}\label{trace}
2\theta^{\prime}_{\alpha^{\prime}\beta^{\prime}}{\cal{W}}_0(x,x^{\prime}) + {\cal{S}}^{\prime}\theta^{\prime}_{\alpha^{\prime}}\cdot{\cal{W}}_{1\beta^{\prime}}(x,x^{\prime})\hspace{1cm} \nn\\
+H^{-2} {D^{\prime}_2}_{\alpha^{\prime}}\bar\partial\cdot{\cal{W}}_{g\beta^{\prime}}(x,x^{\prime})=0,
\end{eqnarray}
or its equivalent form,
\begin{eqnarray}
2\theta_{\alpha\beta}{\cal{W}}_0(x,x^{\prime}) + {\cal{S}}\theta_{\alpha}\cdot{\cal{W}}_{1\beta}(x,x^{\prime})\hspace{1cm} \nn\\
+H^{-2} {D_2}_{\alpha}\bar\partial^\prime\cdot{\cal{W}}_{g\beta}(x,x^{\prime})=0.
\end{eqnarray}
Note that, for the sake of simplicity, we also impose the divergencelessness condition on ${\cal W}_1$.}
\end{itemize}

By imposing the bitensor (\ref{2point 00}) to verify Eq. (\ref{cCasimcount}) with respect to variables $x$ and $x^\prime$ (in the following, we first consider the choice $x$) and using the identities given in Appendix (\ref{Some useful relations}), we obtain\footnote{Note that, we here follow a similar procedure to what we have done in Ref. \cite{BambaII} in order to solve the field equation (\ref{cCasimcount}).}
\begin{eqnarray} \label{W_0}
(Q_0 + 6)\theta^{\prime}{\cal{W}}_0 = -4{\cal{S}}^\prime\theta^\prime\cdot{\cal{W}}_1,
\end{eqnarray}
\begin{eqnarray} \label{W_1}
(Q_1 + 2){\cal{W}}_1 = 0,
\end{eqnarray}
and
\begin{eqnarray} \label{W_g}
(1-c)(Q_1 + 6)D^\prime_2{\cal{W}}_g = (2-5c)H^2{\cal{S}}^\prime x\cdot\theta^\prime{\cal{W}}_1\hspace{0.5cm}\nonumber\\
+ c \Big(-\frac{1}{2} H^2 D_1\theta^\prime{\cal{W}}_0 - T{\cal{S}}^\prime\theta^\prime\cdot\bar\partial{\cal{W}}_1\Big) + D^\prime_2 \Xi_g,
\end{eqnarray}
where $ T {\cal{S}}^\prime\theta^\prime\cdot\bar\partial {\cal{W}}_1 = {\cal{S}}^\prime\theta^\prime\cdot\bar\partial {\cal{W}}_1 - H^2{\cal{S}}^\prime x\theta^\prime\cdot{\cal{W}}_1$ and $D^\prime_2\Xi_g$ appears here because of the canceling property of $D_2$, i.e. $ D_2 (D^\prime_2 \Xi_g)=0$.

Eq. (\ref{W_0}) allows us to express the biscalar ${\cal W}_0$ in terms of ${\cal W}_1$ as follows
\begin{eqnarray}\label{W_0 = 2/3W_1}
\theta^{\prime}{\cal{W}}_0 = -\frac{2}{3} {\cal{S}}^\prime\theta^\prime\cdot{\cal{W}}_1.
\end{eqnarray}
As a matter of fact, considering (\ref{W_1}) supplemented with the divergencelessness condition ($\bar\partial\cdot{\cal{W}}_1=0$)\footnote{The bivector ${\cal{W}}_1$ is transverse, therefore we have $\partial\cdot{\cal{W}}_1=\bar\partial\cdot{\cal{W}}_1$.} results in $Q_0{\cal{W}}_1=0$, which means that
\begin{eqnarray}
Q_0\theta^{\prime}{\cal{W}}_0 = -\frac{2}{3}Q_0 {\cal{S}}^\prime\theta^\prime\cdot{\cal{W}}_1=0.\nonumber
\end{eqnarray}
On the other hand, by decomposing $D^\prime_2{\cal{W}}_g =D^\prime_2 W_g +D^\prime_2 W_g^\Xi$, while
\begin{eqnarray}
(1-c)(Q_1+6)D^\prime_2 W_g^\Xi = D^\prime_2\Xi_g,\;\; x\cdot W_g^\Xi = 0, \; \bar\partial\cdot W_g^\Xi = 0,\nonumber
\end{eqnarray}
one can rewrite the inhomogeneous Eq. (\ref{W_g}) in the following form
\begin{eqnarray}\label{29}
(Q_1+6)D^\prime_2 W_g = \frac{1}{1-c}[-c, \frac{c}{3}, 2-5c],
\end{eqnarray}
where $[u,v,w]\in E$; $E$ is the three-dimensional space constructed over the following linear combination of three basic functions,
\begin{eqnarray}
[u,v,w]=u T{\cal{S}}^\prime\theta^\prime\cdot\bar\partial{\cal{W}}_1 + v H^2{\cal{S}}^\prime D_1\theta^\prime\cdot{\cal{W}}_1 \hspace{1.5cm}\nonumber\\
+ w H^2 {\cal{S}}^\prime x\cdot\theta^\prime{\cal{W}}_1.\nonumber
\end{eqnarray}
These three functions interestingly form a closed set under the action of $(Q_1 +6)$,
\begin{eqnarray}\label{30}
(Q_1+6)T{\cal{S}}^\prime\theta^\prime\cdot\bar\partial{\cal{W}}_1 = [6,2,0],
\end{eqnarray}
\begin{eqnarray}\label{31}
(Q_1+6)H^2{\cal{S}}^\prime D_1\theta^\prime\cdot{\cal{W}}_1 = [0,6,0],
\end{eqnarray}
\begin{eqnarray}\label{32}
(Q_1+6)H^2{\cal{S}}^\prime x\cdot\theta^\prime{\cal{W}}_1 = [-2,0,0].
\end{eqnarray}
On this basis, we can rewrite Eq. (\ref{29}) in the following convenient form
\begin{eqnarray}\label{33}
\left(\begin{array}{ccc} 6 &  0 & -2 \\ 2 & 6 & 0 \\ 0 & 0 & 0 \\
\end{array}\right) \left(\begin{array}{ccc} X \\ Y \\ Z \\ \end{array}\right)= \frac{1}{1-c}\left(\begin{array}{ccc} -c \\ {c}/{3} \\ 2-5c \\ \end{array}\right)
\end{eqnarray}
We here interested in the simplest structure therefore setting $c=2/5=c_\ell$, we get the solution to Eq. (\ref{29}) as
\begin{eqnarray}\label{34}
D^\prime_2 W_g = [0 ,1/27 , 1/3] + \varrho D^\prime_2W_g^\circ,
\end{eqnarray}
where $\varrho$ and $D^\prime_2W_g^\circ$, respectively, stand for an arbitrary constant and a function inside $E$ satisfying
\begin{eqnarray}\label{35}
(Q_1+6)D^\prime_2W_g^\circ= 0.
\end{eqnarray}
It is given, up to a multiplicative constant, by
\begin{eqnarray}\label{36}
D^\prime_2W_g^\circ=[1,-1/3,3].
\end{eqnarray}

In order to make the group theoretical content of our construction more clear, we need to determine the value of $\varrho$. We begin by noting the fact that $D^\prime_2W_g^\circ$ is divergenceless ($D^\prime_2 \bar\partial\cdot W_g^\circ=0$), $D^\prime_2W_g$ however is not, so that we have
\begin{eqnarray}\label{37}
D_1 D^\prime_2 \bar\partial\cdot W_g= [0,1/3,0].
\end{eqnarray}
Considering (\ref{W_0 = 2/3W_1}), the above result is, by the way, consistent with (\ref{trace}). We should also point out that the function $H^2{\cal{S}}^\prime D_1\theta^\prime\cdot{\cal{W}}_1$ is divergenceless, therefore with respect to Eq. (\ref{31}), we have
\begin{eqnarray}
Q_1 (H^2{\cal{S}}^\prime D_1\theta^\prime\cdot{\cal{W}}_1) = Q_0 (H^2{\cal{S}}^\prime D_1\theta^\prime\cdot{\cal{W}}_1) = 0.\nonumber
\end{eqnarray}
This means that, inside the bitensor (\ref{2point 00}), the $H^2 {\cal{S}}^\prime D_2 D_1\theta^\prime\cdot{\cal{W}}_1$ carries the same representation as $\theta\theta^{\prime} {\cal{W}}_0$. On this basis, let us separate $D^\prime_2W_g$ into two parts: $D^\prime_2W_g = D^\prime_2\underline{W}_{g} + D^\prime_2\underline{\underline{{W}}}_{g}$, in which $D^\prime_2\underline{W}_{g}$ is the scalar part of $D^\prime_2W_g$ and $D^\prime_2\underline{\underline{{W}}}_{g}$ is what is left from it.

For $D^\prime_2\underline{\underline{{W}}}_{g}$, considering Eq. (\ref{34}) and Eqs. (\ref{30})-(\ref{32}), we have
\begin{eqnarray}\label{38}
Q_1 D^\prime_2\underline{\underline{{W}}}_{g} = [-2/3-6\varrho, 2\varrho, -2-18\varrho].
\end{eqnarray}
Since the function $D^\prime_2\underline{\underline{{W}}}_{g}$ is not divergenceless, we need to compare Eq. (\ref{38}) with (\ref{37}),
\begin{eqnarray}
D_1 D^\prime_2 \bar\partial\cdot\underline{\underline{{W}}}_{g} = [0, 1/3, 0].
\end{eqnarray}
One can easily observe that $D^\prime_2\underline{\underline{{W}}}_{g}$ verifies
\begin{eqnarray}\label{39}
Q_1 D^\prime_2\underline{\underline{{W}}}_{g} + \frac{2}{3}D_1 D^\prime_2\bar\partial\cdot\underline{\underline{{W}}}_{g} =0,
\end{eqnarray}
if we set
\begin{eqnarray}
\varrho = -1/9.\nonumber
\end{eqnarray}
The group theoretical meaning of this construction is now obvious. Indeed, $D^\prime_2\underline{\underline{{W}}}_{g}$ carries an indecomposable massless representation with spin-1 while the gauge fixing parameter is $c=2/3$ (see \cite{MVF-GGRT}). Accordingly, the solution to Eq. (\ref{W_g}) would be $D^\prime_2{\cal{W}}_g = D^\prime_2\underline{W}_g + D^\prime_2\underline{\underline{{W}}}_{g} + D^\prime_2 W_g^\Xi$, with
\begin{eqnarray}\label{K^g_1}
D^\prime_2\underline{W}_g = [0, 2/27, 0],\;\;\; D^\prime_2\underline{\underline{{W}}}_{g} = [-1/9, 0, 0],
\end{eqnarray}
or equivalently\footnote{For the sake of simplicity, we drop the indices demonstrating $c=c_\ell$ in our notations: ${\cal{W}}^{c=c_\ell} \equiv {\cal{W}}$.}
\begin{eqnarray}\label{2point 00'}
{\cal{W}} = -\frac{2}{3}{\cal{S}}^{\prime} \theta\theta^{\prime}\cdot{\cal{W}}_1 + {\cal{S}}{\cal{S}}^{\prime}\theta\cdot\theta^{\prime}{\cal{W}}_{1}\hspace{2.7cm}\nonumber\\
+ \frac{2}{27} H^2{\cal{S}}^{\prime} D_2D_1 \theta^{\prime}\cdot{\cal{W}}_{1} + D_2 {D^{\prime}_2} (\underline{\underline{{W}}}_{g} + W_g^\Xi).\;\;
\end{eqnarray}

Here and before going any further, it would be interesting to characterize the gauge and the divergencelessness parts of the graviton bitensor two-point function (\ref{2point 00'}). In a totally symmetric way, the gauge part ${D^{\prime}_2} W_g^\Xi$ obeys
\begin{eqnarray} \label{42}
D_2 (Q_1+6) {D^{\prime}_2} W_g^\Xi = 0,
\end{eqnarray}
and considering the identities given in Appendix (\ref{Some useful relations}), we get
\begin{eqnarray}\label{44}
\partial_2\cdot {\cal{W}} = \frac{1}{1-c_\ell} \Big( T{\cal{S}}^{\prime}\theta^\prime\cdot\bar\partial{\cal{W}}_1 - \frac{1}{3} H^2{\cal{S}}^{\prime}D_1\theta^\prime\cdot{\cal{W}}_1 \nonumber\\
+ 3 H^2{\cal{S}}^{\prime} x\cdot\theta^\prime{\cal{W}}_1 \Big).
\end{eqnarray}
The above equation in comparison with (\ref{36}) reveals that
$${D^{\prime}_2}W_g^\circ = ({1-c_\ell}) \partial_2\cdot {\cal{W}}.$$

With respect to (\ref{44}), the graviton bitensor (\ref{2point 00'}) takes the form
\begin{eqnarray} \label{45}
{\cal{W}} &=& \Big( -\frac{2}{3}{\cal{S}}^{\prime} \theta\theta^{\prime}\cdot{\cal{W}}_1 + {\cal{S}}{\cal{S}}^{\prime}\theta\cdot\theta^{\prime}{\cal{W}}_{1} + \frac{1}{3} H^2{\cal{S}}^{\prime} D_2 x\cdot\theta^{\prime}{\cal{W}}_{1} \nonumber\\
&&+ \frac{1}{27} H^2{\cal{S}}^{\prime} D_2D_1 \theta^{\prime}\cdot{\cal{W}}_{1} \Big) +\frac{{c_\ell-1}}{9} D_2\partial_2\cdot {\cal{W}} \nonumber\\
&=& [\mbox{the $c$-independent part}] +\frac{{c_\ell-1}}{9} D_2\partial_2\cdot {\cal{W}}.
\end{eqnarray}
Note that, the gauge part only appears coupled to the scalar part, i.e. $D_2\partial_2\cdot {\cal{W}}$. Here, it should also be noted that the physical graviton two-point function (more precisely, the $c$-independent part) has been already found in the previous papers \cite{II,I}. In the present paper, however, we are interested in the most general form of the graviton two-point function, in $c=c_\ell$ gauge, to construct the graviton quantum field.

Thus far, we have shown that the graviton two-point function can be written in terms of the bivector ${\cal{W}}_1$ satisfying Eq. (\ref{W_1}). Pursuing a similar procedure to what we have done above, we can get the general solution to Eq. (\ref{W_1}) by writting ${\cal{W}}_1$ as the following linear combination of two biscalars ${\cal{W}}_2$ and ${\cal{W}}_3$ \cite{Gazeau5920},
\begin{eqnarray}\label{51}
{\cal{W}}_1 =  \theta\cdot\theta^{\prime} {\cal{W}}_2 + D_1 D^{\prime}_1{\cal{W}}_3.
\end{eqnarray}
Putting the above solution into Eq.(\ref{W_1}), and utilizing the following formulas
\begin{eqnarray}\label{52}
Q_1 D_1 D^{\prime}_1{\cal{W}}_3 = D_1 Q_0 D^{\prime}_1{\cal{W}}_3,
\end{eqnarray}
\begin{eqnarray}\label{53}
Q_1 \theta\cdot\theta^{\prime} {\cal{W}}_2 = \Big(\theta\cdot\theta^{\prime} (Q_0-2) - 2H^2 D_1 x\cdot\theta^{\prime}\Big){\cal{W}}_2,
\end{eqnarray}
we get
\begin{eqnarray}\label{54}
Q_0 {\cal{W}}_2 = 0,
\end{eqnarray}
\begin{eqnarray}\label{55}
D^{\prime}_1 {\cal{W}}_3 = -\frac{1}{2}\Big(\theta^\prime\cdot\bar{\partial} + 2H^2 x\cdot \theta^\prime \Big){\cal{W}}_2.
\end{eqnarray}
Therefore, we can rewrite the general solution (\ref{51}) as
\begin{eqnarray}\label{56}
{\cal{W}}_1 = \Big(\theta\cdot\theta^{\prime} - \frac{1}{2} D_1 [\theta^\prime\cdot\bar{\partial} + 2H^2 x\cdot \theta^\prime] \Big){\cal{W}}_{mc},
\end{eqnarray}
where ${\cal{W}}_2\equiv {\cal{W}}_{mc}$ stands for a minimally coupled biscalar two-point function.

Now by substituting Eq. (\ref{56}) into Eq. (\ref{45}), we can explicitly exhibit the graviton bitensor two-point function (\ref{2point 00}) as the resulting action of a maximally symmetric projection bitensor $\Delta_{\alpha\beta\alpha^{\prime}\beta^{\prime}}(x,x^{\prime})$ on a minimally coupled biscalar two-point function ${\cal{W}}_{mc}(x,x^{\prime})$ as follows
\begin{eqnarray} \label{2point 00''}
{\cal{W}}_{\alpha\beta\alpha^{\prime}\beta^{\prime}}(x,x^{\prime})= \Delta_{\alpha\beta\alpha^{\prime}\beta^{\prime}}(x,x^{\prime}) {\cal{W}}_{mc}(x,x^{\prime}).
\end{eqnarray}

On the other hand, the bitensor (\ref{2point 00}) must verify Eq. (\ref{cCasimcount}) with respect to variable $x^\prime$, as well. Pursuing the same procedure, we obtain
\begin{eqnarray}\label{'}
(Q^{\prime}_0 + 6)\theta{\cal{W}}_0 = -4{\cal{S}}\theta\cdot{\cal{W}}_1,
\end{eqnarray}
\begin{eqnarray}\label{''}
(Q^{\prime}_1 + 2){\cal{W}}_1 = 0,
\end{eqnarray}
\begin{eqnarray}\label{'''}
(1-c)(Q^{\prime}_1 + 6)D_2{\cal{W}}_g = (2-5c)H^2{\cal{S}} x^\prime\cdot\theta{\cal{W}}_1 \hspace{0.5cm} \nonumber\\
+ c \Big(-\frac{1}{2} H^2 D^{\prime}_1\theta{\cal{W}}_0 - T^{\prime}{\cal{S}}\theta\cdot\bar\partial^{\prime}{\cal{W}}_1\Big) + D_2 \Xi_g,
\end{eqnarray}
with $ T^{\prime}{\cal{S}} \theta\cdot\bar\partial^{\prime}{\cal{W}}_1 = {\cal{S}}\theta\cdot\bar\partial^{\prime}{\cal{W}}_1 - {\cal{S}} H^2x^{\prime}\theta\cdot{\cal{W}}_1$ and $D_2\Xi_g$ as an arbitrary bivector satisfies $D^{\prime}_2 (D_2 \Xi_g)=0$.

The solution to Eq. (\ref{'}) can be simply given by
\begin{eqnarray}
\theta{\cal{W}}_0 = -\frac{2}{3} {\cal{S}}\theta\cdot{\cal{W}}_1.\nonumber
\end{eqnarray}
Now with respect to (\ref{''}) along with $\bar\partial^\prime\cdot{\cal{W}}_1=0$, we have $Q^\prime_0{\cal{W}}_1=0$, which implies that
\begin{eqnarray}
Q^\prime_0\theta{\cal{W}}_0 = -\frac{2}{3}Q^\prime_0 {\cal{S}}\theta\cdot{\cal{W}}_1=0.\nonumber
\end{eqnarray}
Regarding the decomposition of $D_2{\cal{W}}_g = D_2W_g + D_2W_g^\Xi$, while
\begin{eqnarray}
(1-c)(Q^\prime_1+6) D_2W_g^\Xi = D_2\Xi_g,\;\; x^\prime\cdot W_g^\Xi = 0,\; \bar\partial^\prime\cdot W_g^\Xi = 0,\nonumber
\end{eqnarray}
Eq. (\ref{'''}) can be written as
\begin{eqnarray}
(Q^\prime_1+6)D_2 W_g = \frac{1}{1-c}\{-c, \frac{c}{3}, 2-5c\}.\nonumber
\end{eqnarray}
The closed three-dimensional space $\{u^\prime,v^\prime,w^\prime\}\in E^\prime$, under the action of $(Q_1^\prime +6)$, is defined by
\begin{eqnarray}
\{u^\prime,v^\prime,w^\prime\} = u^\prime T^\prime{\cal{S}}\theta\cdot\bar\partial^\prime{\cal{W}}_1 + v^\prime H^2{\cal{S}}D^\prime_1\theta\cdot{\cal{W}}_1\hspace{1.5cm}\nonumber\\
 + w^\prime H^2{\cal{S}} x^\prime\cdot\theta{\cal{W}}_1,\nonumber
\end{eqnarray}
while we have
\begin{eqnarray}
(Q^\prime_1+6)T^\prime{\cal{S}}\theta\cdot\bar\partial^\prime{\cal{W}}_1 = \{6,2,0\},\nonumber
\end{eqnarray}
\begin{eqnarray}
(Q^\prime_1+6)H^2{\cal{S}}D^\prime_1\theta\cdot{\cal{W}}_1 = \{0,6,0\},\nonumber
\end{eqnarray}
\begin{eqnarray}
(Q^\prime_1+6)H^2{\cal{S}} x^\prime\cdot\theta{\cal{W}}_1 = \{-2,0,0\}.\nonumber
\end{eqnarray}
Eq. (\ref{'''}) now can be expressed as
\begin{eqnarray}
\left(\begin{array}{ccc} 6 &  0 & -2 \\ 2 & 6 & 0 \\ 0 & 0 & 0 \\
\end{array}\right) \left(\begin{array}{ccc} X^\prime \\ Y^\prime \\ Z^\prime \\ \end{array}\right)=\frac{1}{1-c}\left(\begin{array}{ccc} -c \\ c/3 \\ 2-5c \\ \end{array}\right)
\end{eqnarray}
and the solution to it, setting $c=2/5=c_\ell$, would be
\begin{eqnarray}
D_2 W_g = \{0 ,1/27 , 1/3\} + \varrho^\prime D_2W_g^\circ,\nonumber
\end{eqnarray}
in which $\varrho^\prime$ and $D_2W_g^\circ$ are, respectively, an arbitrary constant and a function inside $E^\prime$, with
\begin{eqnarray}
(Q^\prime_1+6)D_2W_g^\circ= 0,\nonumber
\end{eqnarray}
and therefore
\begin{eqnarray}
D_2W_g^\circ = \{1,-1/3,3\}.\nonumber
\end{eqnarray}

Considering the above calculations, one can easily show that for $\varrho^\prime = - 1/9$, the group theoretical content of this construction becomes clear, so that, the solution to Eq. (\ref{'''}) can be written as $D_2{\cal{W}}_g = D_2\underline{W}_g + D_2\underline{\underline{{W}}}_{g} + D_2 W_g^\Xi$, with
\begin{eqnarray}
D_2\underline{W}_g = \{0, 2/27, 0\},\;\;\; D_2\underline{\underline{{W}}}_{g} = \{-1/9, 0, 0\},\nonumber
\end{eqnarray}
or
\begin{eqnarray}
{\cal{W}} = \theta^{\prime}\theta{\cal{W}}_0 + {\cal{S}}^{\prime}{\cal{S}}\theta^{\prime}\cdot\theta{\cal{W}}_{1}\hspace{4cm}\nonumber\\
+ \frac{2}{27} H^2{\cal{S}} D^{\prime}_2D^{\prime}_1 \theta\cdot{\cal{W}}_{1} + D^{\prime}_2 {D_2} (\underline{\underline{{W}}}_{g} + W_g^\Xi).\nonumber
\end{eqnarray}
Note that, $D_2\underline{W}_g$ is the scalar part of $D_2 W_g$ and $D_2\underline{\underline{{W}}}_{g}$ satisfies
\begin{eqnarray}
Q^\prime_1 D_2\underline{\underline{{W}}}_{g} + \frac{2}{3}D^\prime_1D_2 \bar\partial^\prime\cdot\underline{\underline{{W}}}_{g} =0,\nonumber
\end{eqnarray}
while the gauge and the scalar parts, respectively, obey
\begin{eqnarray}
D^\prime_2 (Q^\prime_1+6) {D_2} W_g^\Xi = 0,\nonumber
\end{eqnarray}
\begin{eqnarray}
\partial^\prime_2\cdot {\cal{W}} = \frac{1}{1-c_\ell} \Big( T^\prime{\cal{S}}\theta\cdot\bar\partial^\prime{\cal{W}}_1 - \frac{1}{3} H^2{\cal{S}}D^\prime_1\theta\cdot{\cal{W}}_1 \nonumber\\
+ 3 H^2 {\cal{S}}x^\prime\cdot\theta{\cal{W}}_1 \Big).\nonumber
\end{eqnarray}
The above equation reveals that $D_2W_g^\circ = ({1-c_\ell}) \partial^\prime_2\cdot {\cal{W}}$.

According to the above formulas, the general bitensor (\ref{2point 00}), with respect to $x^\prime$, can be expressed as
\begin{eqnarray}\label{aa}
{\cal{W}} &=& \Big(-\frac{2}{3}{\cal{S}}\theta^{\prime}\theta\cdot{\cal{W}}_1 + {\cal{S}}^{\prime}{\cal{S}}\theta^{\prime}\cdot\theta{\cal{W}}_{1} + \frac{1}{3} H^2{\cal{S}} D^{\prime}_2 x^{\prime}\cdot\theta{\cal{W}}_{1} \nonumber\\
&&+ \frac{1}{27} H^2{\cal{S}} D^{\prime}_2D^{\prime}_1 \theta\cdot{\cal{W}}_{1} \Big) +\frac{{c_\ell-1}}{9} D^{\prime}_2\partial^{\prime}_2\cdot {\cal{W}}.
\end{eqnarray}

In (\ref{aa}), the bivector ${\cal{W}}_1$ can be obtained by putting the following linear combination into Eq. (\ref{''})
\begin{eqnarray}
{\cal{W}}_1 =  \theta^{\prime}\cdot\theta {\cal{W}}_2 + D^{\prime}_1 D_1{\cal{W}}_3,
\end{eqnarray}
for which, we have
\begin{eqnarray}
Q^{\prime}_0 {\cal{W}}_2 = 0,\;\; D_1 {\cal{W}}_3 = -\frac{1}{2}\Big(\theta\cdot\bar{\partial}^{\prime} + 2H^2 x^{\prime}\cdot \theta\Big){\cal{W}}_2.
\end{eqnarray}
Therefore, ${\cal{W}}_1$ can be presented as (${\cal{W}}_2\equiv {\cal{W}}_{mc}$)
\begin{eqnarray}\label{bb}
{\cal{W}}_1 = \Big(\theta^{\prime}\cdot\theta - \frac{1}{2} D^{\prime}_1 [\theta\cdot\bar{\partial}^{\prime} + 2H^2 x^{\prime}\cdot\theta] \Big){\cal{W}}_{mc}.
\end{eqnarray}

Again, by combining Eqs. (\ref{aa}) and (\ref{bb}) together, we can demonstrate the graviton two-point function as the action of a maximally symmetric projection bitensor on a minimally coupled biscalar two-point function
\begin{eqnarray}\label{2point 00''new}
{\cal{W}}_{\alpha\beta\alpha^{\prime}\beta^{\prime}}(x,x^{\prime})= \Delta^{\prime}_{\alpha\beta\alpha^{\prime}\beta^{\prime}}(x,x^{\prime}) {\cal{W}}_{mc}(x,x^{\prime}).
\end{eqnarray}

To summarize, thus far we have shown that the graviton bitensor two-point function (\ref{2point 00}), with respect to $x$ and $x^\prime$, can be written in terms of a maximally symmetric projection bitensor on a minimally coupled biscalar two-point function (see (\ref{2point 00''}) and (\ref{2point 00''new})). Our result, therefore, would be dS invariant if and only if its structure function, the minimally coupled biscalar two-point function ${\cal W}_{mc}$, could be written in a dS-invariant form. Respecting the result given by Allen and Folacci in their seminal work \cite{AllenFolacci} (see also \cite{Gazeau1415,de Bievre6230}), it is clear that if one needs the biscalar massless minimally coupled two-point function to be dS invariant, has to ignore its analyticity properties for the time being \cite{Bertola,BrosPRL}. This means that ${\cal{W}}_{mc}(x,x^{\prime})$ is only a function of the invariant length ${\cal{Z}}= -H^2x\cdot x^{\prime}$; ${\cal W}_{mc} = {\cal W}_{mc}({\cal Z})$. In the same way, the graviton two-point function can be written in a full dS-invariant form only through ignoring its analyticity properties: the only dS-invariant two-point function that naturally appears in the case of the linearized gravity in dS spacetime is nothing but the commutator. This commutator is of course not of positive type (see the previous section).

Now, considering ${\cal W}_{mc} = {\cal W}_{mc}({\cal Z})$ and using the identities given in Appendix (\ref{Some useful relations}), we have
\begin{widetext}
\begin{eqnarray}
\theta^{\prime}_{\alpha^{\prime}\beta^{\prime}}{\cal W}_{0}(x,x^{\prime})=\frac{1}{3}{\cal S}^{\prime} \Big[\theta^{\prime}_{\alpha^{\prime}\beta^{\prime}}
+ \frac{4}{1-{\cal{Z}}^2}H^2 (x\cdot\theta^{\prime}_{\alpha^{\prime}})(x\cdot\theta^{\prime}_{\beta^{\prime}})\Big]{\cal{Z}}\frac{d}{d{\cal{Z}}}{\cal W}_{mc}({\cal{Z}}),
\end{eqnarray}
\begin{eqnarray}
{\cal W}_{1\beta\beta^{\prime}}(x,x^{\prime})=\frac{1}{2} \Big[\frac{3+{\cal{Z}}^2}{1-{\cal{Z}}^2}H^2 (x^{\prime}\cdot\theta_{\beta})(x\cdot\theta^{\prime}_{\beta^{\prime}})
- {\cal{Z}} (\theta_{\beta}\cdot\theta^{\prime}_{\beta^{\prime}})\Big]\frac{d}{d{\cal{Z}}}{\cal W}_{mc}({\cal{Z}}),
\end{eqnarray}
\begin{eqnarray}
D_{2\alpha}D^{\prime}_{2\alpha^{\prime}}{\cal W}_{g\beta\beta^{\prime}}(x,x^{\prime}) =
\frac{H^2}{54(1-{\cal{Z}}^2)^2}{\cal S}{\cal S}^{\prime}\Big[ H^{-2}{\cal{Z}}(7 - 3{\cal{Z}}^2)(1-{\cal{Z}}^2) \theta_{\alpha\beta} \theta^{\prime}_{\alpha^{\prime}\beta^{\prime}}
+ 24{\cal{Z}}^3 \theta_{\alpha\beta}(x\cdot\theta^{\prime}_{\alpha^{\prime}})(x\cdot\theta^{\prime}_{\beta^{\prime}})\nonumber\\
+ {\cal{Z}}( -1 + 9{\cal{Z}}^2 )(1-{\cal{Z}}^2)H^{-2}(\theta_{\alpha}\cdot\theta^{\prime}_{\alpha^{\prime}}) (\theta_{\beta}\cdot\theta^{\prime}_{\beta^{\prime}})
+ ( -1 - 50{\cal{Z}}^2 -45{\cal{Z}}^4 )(\theta_{\alpha}\cdot\theta^{\prime}_{\alpha^{\prime}}) (x\cdot\theta^{\prime}_{\beta^{\prime}}) (x^{\prime}\cdot\theta_{\beta})\nonumber\\
+ \frac{36{\cal{Z}}+120{\cal{Z}}^3+36{\cal{Z}}^5}{1-{\cal{Z}}^2}
H^2(x^{\prime}\cdot\theta_{\alpha})(x^{\prime}\cdot\theta_{\beta})(x\cdot\theta^{\prime}_{\alpha^{\prime}})(x\cdot\theta^{\prime}_{\beta^{\prime}})
+ 12{\cal{Z}}(3-{\cal{Z}}^2)\theta^{\prime}_{\alpha^{\prime}\beta^{\prime}}(x^{\prime}\cdot\theta_{\alpha})(x^{\prime}\cdot\theta_{\beta})\Big] \frac{d}{d{\cal{Z}}}{\cal W}_{mc}({\cal{Z}}).
\end{eqnarray}
According to the above formulas, the explicit form of the graviton commutator can be written as follows
\begin{eqnarray}\label{WMMC}
{\cal W}_{\alpha\beta \alpha^{\prime}\beta^{\prime}}(x,x^{\prime}) = {\frac{2{\cal{Z}}}{27(1-{\cal{Z}}^2)^2}}{\cal S}{\cal S}^{\prime} \Big[(\theta_{\alpha}\cdot\theta^{\prime}_{\alpha^{\prime}})(\theta_{\beta}\cdot\theta^{\prime}_{\beta^{\prime}})f_1({\cal{Z}})
+ H^2\Big(\theta^{\prime}_{\alpha^{\prime}\beta^{\prime}}(x^{\prime} \cdot \theta_{\alpha}) (x^{\prime}\cdot\theta_{\beta})+\theta_{\alpha\beta}(x\cdot\theta^{\prime}_{\alpha^{\prime}})(x\cdot\theta^{\prime}_{\beta^{\prime}}) \Big)f_2({\cal{Z}})\nonumber\\
+\theta_{\alpha\beta}\theta^{\prime}_{\alpha^{\prime}\beta^{\prime}}f_3({\cal{Z}})
+H^4(x^{\prime}\cdot\theta_{\alpha})(x^{\prime}\cdot\theta_{\beta})(x\cdot\theta^{\prime}_{\alpha^{\prime}})(x\cdot\theta^{\prime}_{\beta^{\prime}}) f_4({\cal{Z}})
+(\theta_{\alpha}\cdot\theta^{\prime}_{\alpha^{\prime}})(x\cdot\theta^{\prime}_{\beta^{\prime}})(x^{\prime}\cdot\theta_{\beta})f_5({\cal{Z}}) \Big]\frac{d}{d{\cal{Z}}}{\cal W}_{mc}({\cal{Z}}),\;\;\;
\end{eqnarray}
\end{widetext}
where
\begin{eqnarray}
f_1({\cal{Z}}) &=& (1-{\cal{Z}}^2) (-7 + 9{{\cal{Z}}}^2),\nonumber\\
f_2({\cal{Z}}) &=& 3(3 - {{\cal{Z}}}^2),\nonumber\\
f_3({\cal Z})&=&\frac{1}{4}(16-3{\cal Z}^2) (1-{{\cal{Z}}}^2),\nonumber\\
f_4({\cal{Z}}) &=& \frac{3}{(1-{\cal Z}^2)}(3 + 10 {{\cal{Z}}}^2 + 3{{\cal{Z}}}^4),\nonumber\\
f_5({\cal{Z}}) &=& \frac{1}{{2\cal{Z}}} (41 - 50{\cal{Z}}^2 - 36 {\cal{Z}}^4 ),\nonumber
\end{eqnarray}
and ${\cal W}_{mc}({\cal{Z}})$ obeys
\begin{eqnarray}
Q_0{\cal W}_{mc}({\cal{Z}}) = \Big( (1-{\cal{Z}}^2)\frac{d^2}{d{\cal{Z}}^2} - 4{\cal{Z}} \frac{d}{d{\cal{Z}}} \Big){\cal{W}}_{mc}({\cal{Z}}) = 0.\nonumber
\end{eqnarray}
The general solution to the above equation is
\begin{eqnarray}
{\cal{W}}_{mc}({\cal{Z}}) = C_1 \Big( \frac{-2{\cal Z}}{1- {\cal{Z}}^2} + \ln(\frac{1-\cal{Z}}{1+{\cal{Z}}}) \Big) + C_2,\nonumber
\end{eqnarray}
where $C_1,C_2\in\mathbb{R}$.

Using the formula
\begin{eqnarray}
\ln(x\pm i0) &=& \ln(|x|) \pm i\pi\theta(-x),\nonumber
\end{eqnarray}
it is easily seen that $\ln(\frac{1-\cal{Z}}{1+{\cal{Z}}})$ is a non-local function \cite{Bertola}. However, in the commutator (\ref{WMMC}), ${\cal{W}}_{mc}({\cal{Z}})$ enters only via its derivative,
\begin{equation} \label{WCMMC2'}
\frac{d}{d{\cal{Z}}}{\cal{W}}_{mc}({\cal{Z}}) = -\frac{4C_1}{({\cal{Z}}^2 - 1)^2},
\end{equation}
which is a local function. One can now easily see that for points that are not null-related our obtained result (\ref{WMMC}) is finite.

Here, it would be also convenient to demonstrate our result in the intrinsic coordinate (see Appendix (\ref{bitensors})),
\begin{eqnarray}\label{WMMCI}
{\cal{W}}_{\mu\nu\mu'\nu'}(X,X') =  \frac{2{\cal{Z}}}{27}{\cal{S}}{\cal{S}}'\Big[\frac{f_1}{(1-{\cal{Z}}^2)^2}g_{\mu\mu'}g'_{\nu\nu'}\hspace{1.3cm}\nonumber\\
+ \frac{f_2}{1-{\cal{Z}}^2}(g_{\mu\nu}n_{\mu'}n_{\nu'}+g'_{\mu'\nu'}n_\mu n_\nu)
+ \frac{f_3}{(1-{\cal{Z}}^2)^2}g_{\mu\nu}g'_{\mu'\nu'}\nonumber\\
+ \Big(\frac{f_1}{(1+{\cal{Z}})^2} + f_4-\frac{f_5}{1+{\cal{Z}}}\Big)n_\mu n_\nu n_{\mu'}n_{\nu'}\nonumber\\
+ \Big(\frac{2({\cal{Z}}-1)f_1}{(1-{\cal{Z}}^2)^2}+\frac{f_5}{1-{\cal{Z}}^2} \Big)g_{\mu\mu'}n_\nu n_{\nu'}\Big]\frac{d}{d{\cal{Z}}}{\cal{W}}_{mc}({\cal{Z}}).\;\;\;
\end{eqnarray}

We end our discussion in this section by commenting on some existing results which are in contradiction with ours. In Refs. \cite{higuchi1,HH Frob} the graviton two-point function in dS spacetime has been given in terms of maximally symmetric bitensors. More precisely, it has been argued that without ignoring the analyticity properties for the time being the dS-invariant, infrared finite graviton two-point function of positive type is quite achievable (see the same argument in \cite{Bernar,HH Higuchi'',HH Bernar,Fewster,Faizal&Higuchi,HH Higuchi',Marolf&Morrison,HH Higuchi,HH Hartle,HH Allen,HH Marolf,FrobJCAP}). Take a closer look at the method which has been utilized in Refs. \cite{higuchi1,HH Frob,Bernar,HH Higuchi'',HH Bernar,Fewster,Faizal&Higuchi,HH Higuchi',Marolf&Morrison,HH Higuchi,HH Hartle,HH Allen,HH Marolf,FrobJCAP} shows that the authors have considered the synchronous-transverse-traceless (STT) gauge to evaluate the graviton two-point function. The critical point associated with this method is that it is not possible to find a dS graviton field satisfying the STT gauge conditions if $L=0$ or $1$ \cite{Higuchizzz3}. Indeed, by solving the field equation in the conformal coordinate (\ref{Linequ}) directly, the normalization factor associated with the lowest graviton eigenmode, i.e. $L=0$, would be broken \cite{AHigu1, AHigu2, AHigu3}. At first glance, this mode can be regarded as a spurious mode, and thus can be dropped from the mode expansion of the graviton field. It is also discussed that the mode $L=1$ suffers from an analogous difficulty (see the details in the references cited above). However, as already pointed out, thanks to the mathematical structure of the formulation we are using, inspired by the ambient space approach, it can be checked quite easily that the aforementioned break down in the normalization factor is because of a degeneracy for $L=0$ mode reflecting the gauge-like symmetry (\ref{gauge-like}). This point explicitly reveals that, quite contrary to the authors claim in \cite{higuchi1,HH Frob,Bernar,HH Higuchi'',HH Bernar,Fewster,Faizal&Higuchi,HH Higuchi',Marolf&Morrison,HH Higuchi,HH Hartle,HH Allen,HH Marolf,FrobJCAP}, such a construction in the STT gauge conditions ($L\geq2$), by ignoring the $L=0$ mode and consequently the gauge-like symmetry (\ref{gauge-like}) reflected by it, does not transform correctly under the whole symmetries of the classical theory, and therefore, even obtaining an infrared free graviton two-point function in this way is not physically significant since it is not covariant anyway. To see other criticism to this method, one can refer to \cite{111,222,333,444,555}.

\section{De Sitter linear quantum gravity}

\subsection{The quantum field}
The explicit knowledge of the commutator ${\cal W}_{\mu\nu\mu'\nu'}(X,X')$, given in the light-cone gauge ($c=c_\ell$), with the above-mentioned properties allows us to construct an acceptable quantum theory of the dS graviton field. The fields $\underline{{\cal K}}_{\mu\nu}(X)$, which we wish to consider on $M_H$, are expected to be operator-valued distributions on $M_H$ acting on the bosonic Fock space $\underline{{\cal H}}\oplus \underline{{\cal H}}^\ast$ built on the total (Krein) space ${\cal H}\oplus {\cal H}^\ast$. In terms of Fock space and field operator, the aforementioned properties of ${\cal W}_{\mu\nu\mu'\nu'}(X,X')$ become equivalent to the following conditions:
\begin{itemize}
\item{Existence of an indecomposable representation of the dS group $\underline{U}_{\cal G}$ which is the extension of the dS natural representation $U_{\cal G}$ (see (\ref{complete})) on the space of states to the Fock space.}
\item{Existence of a distinguished vector $|\Omega\rangle$, called ``the vacuum", in the Fock space, cyclic for the polynomial algebra of field
operators and invariant under the representation $\underline{U}_{\cal G}$ of the dS group.}
\item{Existence of a complex vector space ${\cal A}$, with an indefinite sesquilinear form, that can be described as the direct sum
\begin{eqnarray}
{\cal A} = {\cal A}_0\oplus [\oplus_n {\cal S}({\cal A}_1)^{\otimes n}],\nonumber
\end{eqnarray}
in which ${\cal A}_0= \{ \vartheta |\Omega\rangle, \vartheta\in\mathds{C} \}$, and ${\cal A}_1$ is defined with the indefinite sesquilinear form.}
\item{Covariance of the field operators under the representation $\underline{U}_{\cal G}$, i.e.,
\begin{eqnarray}\label{sfomodes}
\underline{U}_{\cal G}\underline{{\cal K}}(X)\underline{U}_{{\cal G}^{-1}} = \underline{{\cal K}}({\cal G}X),\nonumber
\end{eqnarray}
for any ${\cal G}$ in the dS group.}
\item{Locality,
$$[\underline{{\cal K}}(X),\underline{{\cal K}}(X')]=0,$$
as long as the points $X$ and $X'$ are not causally connected.}
\item{Transversality, $x\cdot\underline{{\cal K}} = 0$.}
\item{Index symmetrizer, $\underline{{\cal K}}_{\mu\nu}=\underline{{\cal K}}_{\nu\mu}$.}
\item{Tracelessness, $\underline{{\cal K}}^\prime = 0$.}
\end{itemize}

Let us now define the graviton quantum field corresponding to the above commutator. For any test function $f_{\mu\nu}\in {\cal D}(M_H)$, we define the vector valued distribution taking values in the space $V^c = \mbox{span} \{{\cal K}^j, {\cal K}^{j\ast}\}$, with $j\in {\cal J}$,\footnote{Here, for the sake of simplicity, we use the notation $${\cal J}=\{(L,l,m)\in \mathbb{N}\times\mathbb{N}\times\mathbb{Z};\; L\geq 0, 0\leq l\leq L, -l\leq m\leq l\}.$$} by
\begin{eqnarray}\label{uniqele}
X\rightarrow p_{\mu\nu}(f)(X) &=& \int_{M_H}{\cal W}_{\mu\nu\mu'\nu'}(X,X')f^{\mu'\nu'}(X')d\sigma(X') \nonumber\\
&=& \sum_j{\cal K}^j(f){\cal K}^j_{\mu\nu}(X).
\end{eqnarray}
Here, ${\cal K}^j(f)$ is defined by
\begin{eqnarray}\label{sfomodes}
{\cal K}^j(f) = \int_{M_H}{\cal K}^{j\ast}_{\mu\nu}(X)f^{\mu\nu}(X)d\sigma (X).
\end{eqnarray}
It is indeed the smeared form of the modes. The space generated by the vector valued distributions ($p(f)$'s) is equipped with the indefinite invariant inner product ($\forall f,g\in {\cal D}(M_H)$)
\begin{eqnarray}\label{eqiii}
\langle p(f), p(g)\rangle = \int_{M_H\times M_H}f^{\ast \; \mu\nu}(X){\cal W}_{\mu\nu\mu'\nu'}(X,X')\nonumber\\
\times g^{\mu'\nu'}(X')d\sigma(X')d\sigma(X).
\end{eqnarray}

As usual, the field is expected to be an operator-valued distribution
\begin{eqnarray}\label{opvadi}
\underline{{\cal K}}(f) = a(p(f)) + a^\dagger(p(f)),
\end{eqnarray}
or in the unsmeared form
\begin{eqnarray}\label{unsopvadi}
\underline{{\cal K}}(X) = a(p(X)) + a^\dagger(p(X)).
\end{eqnarray}
Here, the operators $a$ and $a^\dagger$ are, respectively, antilinear and linear in the argument. Hence, the field reads
\begin{eqnarray}\label{qf}
\underline{\cal{K}}(X) &=&  \sum_{j\in {\cal J}} (a_j {\cal{K}}^{j}(X) + h.c.)\nonumber\\
&& - \sum_{j\in {\cal J}} (b_j {\cal{K}}^{j\ast} (X) + h.c.),
\end{eqnarray}
where the operators $a_j$ and $b_j$ are, respectively, the annihilators of the modes ${\cal K}^j$ and ${\cal K}^{j\ast}$. The non-vanishing commutation relations between these operators are
\begin{eqnarray}\label{commutre}
[a_j, a_{j^\prime}^\dagger] = \delta_{jj^\prime} = -[b_j, b_{j^\prime}^\dagger].
\end{eqnarray}
These operators are defined by
\begin{eqnarray}\label{vacuum}
a_j\mid\Omega\rangle = 0 = b_j\mid\Omega\rangle,
\end{eqnarray}
where $\mid\Omega\rangle$ is a dS-invariant vacuum. We call it the Krein-Gupta-Bleuler (KGB) vacuum.

Finally, one finds the commutation relations between fields as follows
\begin{eqnarray}\label{cauoffield}
[\underline{{\cal K}}(X),\underline{{\cal K}}(X')] & = & 2i\mbox{Im}\langle p(X), p(X')\rangle \nonumber\\
& = & 2i\mbox{Im}{\cal W}(X,X').
\end{eqnarray}

Having shown that the field we constructed is causal and has all the covariance properties of the classical field, we can now turn to an investigation of the physical content of the theory. In this regard, in the following part, to interpret the theory, we define its physical states and its observables. We particularly discuss that the presence of non-physical states do not yield any trouble in the theory (e.g. the appearance of negative energies).

\subsection{Quantum observables}
Let us start with determining the physical states. Considering the chain (\ref{trip2}), the whole Fock space $\underline{V}^c = \underline{{\cal H}}\oplus \underline{{\cal H}}^\ast$ has the following second-quantized Gupta-Bleuler structure
\begin{eqnarray}\label{focktrip1}
\underline{\underline{V}}_{g_1}\subset \underline{V}_1\subset \underline{V}^c.
\end{eqnarray}
Here, we designate by $\underline{V}_1$ the space generated from the Fock vacuum by creating elements of $V_1$. It is actually the space generated by\footnote{By $j\in {\cal J}\geq1$, in comparison with $j\in {\cal J}$, we mean the set of $\{(L,l,m)\}$, with $L\geq1$.}
\begin{eqnarray}
\underline{V}_1 \equiv \{(a^\dagger_{(g_1)})^{n^{}_{0}} \prod_{j\in{\cal{J}}\geq 1} (a^\dagger_{j})^{n^{}_j} |\Omega\rangle \},\nonumber
\end{eqnarray}
where $a^\dagger_{(g_1)}\equiv a^\dagger ({\cal K}_{g_1})$. We also designate by $\underline{\underline{V}}_{g_1}$ the subspace of gauge-like states that are orthogonal to $\underline{V}_1$,
\begin{eqnarray}
\Psi\in\underline{\underline{V}}_{g_1}\;\;\;\;\mbox{iff}\;\;\;\;\Psi\in\underline{V}_1
\;\;\;\;\mbox{and}\;\;\;\;\langle\Psi,\Phi\rangle=0,\;\;\;\;\forall\Phi\in\underline{V}_1.\nonumber
\end{eqnarray}
Note that the subspace $\underline{\underline{V}}_{g_1}$ is strictly greater than $\underline{V}_{g_1}$ defined by
\begin{eqnarray}
\underline{V}_{g_1} \equiv \{(a^\dagger_{(g_1)})^{n^{}_{0}} |\Omega\rangle \}.\nonumber
\end{eqnarray}
Indeed, for any state $\Psi \in \underline{V}_1$, the state $(a^\dagger_{(g_1)})^{n^{}_{0}}\Psi$ belongs to $\underline{\underline{V}}_{g_1}$ but not to $\underline{V}_{g_1}$.

In the above structure, the quotient space $\underline{V}_1/\underline{\underline{V}}_{g_1}$ contains all physical states, but it is not restricted to them: with respect to the evident gauge symmetry, it also contains some non-physical states. Indeed, as already pointed out, in the case of the dS linearized quantum gravity, we encounter with two kinds of the gauge degrees of freedom which are distinguished by their Gupta-Bleuler structures. Technically, according to the evident gauge symmetry, another Gupta-Bleuler structure also appears in the Fock space
\begin{eqnarray}\label{focktrip1}
\underline{\underline{V}}_{g_2}\subset \underline{V}_2\subset \underline{V}^c,
\end{eqnarray}
where
\begin{eqnarray}
\underline{V}_2 \equiv \{ \prod_{j\in{\cal{J}}}(\textbf{a}^\dagger_{(g_2)j})^{m_j}(\textbf{b}^\dagger_{(g_2)j})^{m^\prime_j}   (\textbf{a}^\dagger_{(\frac{2}{g_2})j})^{n^{}_{j}}(\textbf{b}^\dagger_{(\frac{2}{g_2})j})^{n^\prime_j} |\Omega\rangle\},\nonumber
\end{eqnarray}
with
$$ \textbf{a}^\dagger_{(g_2)j} \equiv a^\dagger ({\cal K}^j),\;\; \textbf{b}^\dagger_{(g_2)j} \equiv b^\dagger ({\cal K}^{j\ast}),\;\; \forall {\cal K}^j, {\cal K}^{j\ast}\in V_{g_2},$$
and
$$\textbf{a}^\dagger_{(\frac{2}{g_2})j} \equiv a^\dagger ({\cal K}^{j}),\;\; \textbf{b}^\dagger_{(\frac{2}{g_2})j} \equiv b^\dagger ({\cal K}^{j\ast}), \;\; \forall {\cal K}^j, {\cal K}^{j\ast}\in V_2/V_{g_2}.$$
Here, $\underline{\underline{V}}_{g_2}$ is the space of gauge states which is orthogonal to $\underline{V}_2$.

The physical states, therefore, would be those elements of $\underline{V}_1/\underline{\underline{V}}_{g_1}$ which can also be found in the quotient space $\underline{V}_2/\underline{\underline{V}}_{g_2}$. More accurately, they are the elements of the invariant subspace
\begin{eqnarray}
{\underline{V}}_{phys} &=& \Big(\underline{V}_1/\underline{\underline{V}}_{g_1}\Big)\bigcap \Big(\underline{V}_2/\underline{\underline{V}}_{g_2}\Big) \nonumber\\
&=& \{\prod_{j\in{\cal{J}}\geq 1} (\textbf{a}^\dagger_{(\frac{2}{g_2})j})^{n^{}_{j}}|\Omega\rangle\}.
\end{eqnarray}
It is also should be noted that two physical states, e.g. $\Psi$ and $\Psi^\prime$, are physically equivalent if they differ by an element of gauge and/or gauge-like states ($\Psi - \Psi^\prime$ belongs to $\underline{\underline{V}}_{g_2}$ and/or $\underline{\underline{V}}_{g_1}$).

Before going to define the observables of the theory, it will be well to clear up one point. According to the Gupta-Bleuler structures associated with the evident gauge symmetry and the gauge-like symmetry, the spaces of the dS-invariant states of $\underline{V}^c$ are $\underline{V}_{g_1}$ and $\underline{V}_{g_2}$, which are, respectively, generated from the vacuum by $(a^\dagger_{(g_1)})^{n^{}_{0}}$ and $(\textbf{a}^\dagger_{(g_2)j})^{m_j}(\textbf{b}^\dagger_{(g_2)j})^{m^\prime_j}$. These spaces are, respectively, infinite-dimensional subspace of $\underline{\underline{V}}_{g_1}$ and $\underline{\underline{V}}_{g_2}$, hence one may say that the Fock vacuum is not the only dS-invariant state. Nevertheless, it is obvious that all these states are physically equivalent to an element of the one-dimensional space generated by the vacuum state. In this sense, we  can say that the Fock vacuum is unique. [Of course, this does not mean that the Bogolyubov transformations, which are indeed changes of physical states, are no longer valid in this construction: the above construction gives a framework in which, instead of having a multiplicity of vacua, we have several possibilities for the space of physical states, while there exist only one field and one vacuum (the latter being invariant under Bogoliubov transformations). See \cite{Garidi2005} for a detailed discussion on the Bogolyubov transformations in the Krein space.]

We are now in a position to define the observables of the theory. As is well-known, observables are defined by the property that they do not ``see" the gauge states: technically, with the assumption that $\Psi$ and $\Psi^\prime$ are equivalent physical states of the system, we must have
\begin{eqnarray}\label{obphyeq}
\langle \Psi |A|\Psi\rangle = \langle \Psi^\prime |A|\Psi^\prime\rangle,
\end{eqnarray}
for any observable $A$ which is a symmetric operator on the Fock space. We note that, according to the above discussion, the field itself is not an observable.

In this context, let us discuss the general features of the stress tensor $T_{\mu\nu}$ which is primary among the observables. In general, the mean values of the stress tensor on the KGB vacuum state is
\begin{eqnarray}\label{qqq}
\langle \Omega|T_{\mu\nu}|\Omega \rangle &=& \sum_{j\in{\cal{J}}} T_{\mu\nu} [{\cal{K}}_{\mu\nu}^{j},{\cal{K}}_{\mu\nu}^{j\ast}]\nn\\
&& - \sum_{j\in{\cal{J}}}T_{\mu\nu} [{\cal{K}}_{\mu\nu}^{j\ast}, {\cal{K}}_{\mu\nu}^{j}] = 0,
\end{eqnarray}
where $T_{\mu\nu} [{\cal{K}},{\cal{K}}]$ denotes the bilinear expression of the stress tensor $T_{\mu\nu}$. The cancellation appeared in (\ref{qqq}), because of the unusual second term with the minus sign, is indeed due to the terms of the quantum field (\ref{qf}) containing $b_j$ and $b_j^\dagger$. As a direct consequence, the vacuum energy of the free field in this construction automatically vanishes, $\langle \Omega|T_{00}|\Omega \rangle = 0$. This property has an interesting link to the cosmological constant problem (see \cite{CCP}).

Similarly, we can evaluate the behavior of the mean values of the stress tensor on physical states $ |\overrightarrow{{\cal{P}}} \rangle\in {\underline{V}}_{phys}$. Obviously, the same cancellation occurs again and we have
\begin{eqnarray}\label{EMT1}
\langle \overrightarrow{{\cal{P}}}| T_{\mu\nu} |\overrightarrow{{\cal{P}}} \rangle = 2 \mbox{Re} \sum_{j\in {\cal J}\geq1} n_j T_{\mu\nu} [{\cal{K}}_{\mu\nu}^{j},{\cal{K}}_{\mu\nu}^{j\ast}].
\end{eqnarray}
As a result, for any physical state $|\overrightarrow{{\cal{P}}}\rangle$, one gets
$$\langle\overrightarrow{{\cal{P}}}|T_{00}|\overrightarrow{{\cal{P}}} \rangle > 0,$$
which along with (\ref{qqq}), despite the use of negative norm solutions in the definition of the field, guarantees the positivity of the energy operator in all physical states. Indeed, this construction remarkably provides a framework allowing for an automatic and covariant renormalization of the stress tensor (note that the above expression is free of any infinite term) which meets the so-called Wald axioms, namely: (\emph{i}) $\langle T_{\mu\nu}\rangle$ is covariant and causal since the field itself is, (\emph{ii}) $\langle \overrightarrow{{\cal{P}}}|T_{00}|\overrightarrow{{\cal{P}}} \rangle \geq 0$, for any physical state $|\overrightarrow{{\cal{P}}}\rangle$ (the equality holds if and only if $|\overrightarrow{{\cal{P}}}\rangle = |\Omega\rangle$), and finally (\emph{iii}) the commutation relation $[b_j,b_j^\dagger] = -1$ implies that
$$ a_ja_j^\dagger + a_j^\dagger a_j +  b_jb_j^\dagger + b_j^\dagger b_j = 2a_j^\dagger a_j + 2b_j^\dagger b_j,$$
which is equivalent to reordering when we compute the mean values of the stress tensor on physical states.

\section{Discussion}
We have presented a new (free) graviton quantum field in dS spacetime which transforms correctly under isometries, gauge transformations, and gauge-like transformations, admitting a dS-invariant vacuum \emph{\`{a} la } Gupta-Bleuler, known as the Krein-Gupta-Bleuler vacuum. Let us recall that if one uses the natural dS vacuum state (the Bunch-Davies vacuum) to construct the graviton quantum field not only dS invariance but also gauge-like invariance of the theory would be broken \cite{Hidden,BambaII,BambaI,111,222,333,444,555}. The KGB quantization approach which does not prohibit negative norm states in the definition of the field, however, provides a unified framework to treat gauge and gauge-like symmetry of the theory. The use of this quantization scheme, as we have shown, is justified by the fact that the theory possesses all the properties one might expect from a free field on dS spacetime with high symmetry, namely:
\begin{itemize}
\item{All physical states have positive norms, as needed for a reasonable quantum mechanical interpretation of the theory (of course, all positive norm states are not physical).}
\item{The graviton field we have constructed is causal and it is covariant in the usual strong sense.}
\item{The graviton field transforms correctly under gauge and gauge-like transformations.}
\item{In spite of the fact that the operator $T_{00}(X)$ is not positively definite as an operator on the full space of states, the expected values of $T_{00}(X)$ between excited physical states are clearly positive.}
\end{itemize}

Beside these, there are at least three related but distinct reasons for which one might be interested in the KGB quantization method:
\begin{itemize}
\item{This method fulfills all common results in the flat limit \cite{Garidi2005}.}
\item{It provides a remarkable automatic and covariant renormalization mechanism of the mean value of the stress tensor verifying the so-called Wald axioms (interestingly, the vacuum energy independent of the curvature is zero) \cite{braneI,braneII,blackhole}.}
\item{In the semiclassical description of general relativity ($\Lambda>0$) when matter field is present, a preliminary estimate of the expected order of magnitude of vacuum energy density stored in the cosmological constant today with respect to the KGB vacuum yields a remarkable coincidence with the empirical data \cite{CCP}.}
\end{itemize}

On this basis, while there is definitely a tremendous amount of work still to be performed, we would argue that the KGB method could very well in its own right be a promising alternative for quantum field theory in the presence of a non-zero (positive) cosmological constant. Last but certainly not least, this method is one of very few choices that has any realistic hope of direct confrontation with the long-standing problem of dS breaking in linearized quantum gravity and the dark energy problem, while it recovers all common results in the Minkowskian limit.

\section*{Acknowledgements}
This work was supported by the National Natural Science Foundation of China with the Grants Nos: 11375153 and 11675145.

\begin{appendix}

\setcounter{equation}{0}
\section{\label{dS wave planes}dS plane waves as generating functions}
The fields carrying the representations $\Pi_{p,0}$, with $p=1,2,..$, in the scalar discrete series are solutions to the following equation (issued from equation (\ref{fieqca}))
\begin{eqnarray}\label{Caopeq}
(Q_0+\kappa(\kappa + 3))\Phi(x)=0,
\end{eqnarray}
where we have considered the unifying complex parameter $\kappa$ given by $\kappa =p-1$ or $=-p-2$. [The irreducible representation associated with our study is the massless minimally coupled representation $\Pi_{1,0}$, the lowest case in the scalar discrete series representations, corresponds to $\kappa = -3$. Note that the case $\kappa =0$ can be considered as the associated trivial representation.] With the notation of (\ref{planew}), a continuous family of simple solutions to (\ref{Caopeq}) can be written as
\begin{eqnarray}\label{dspwave}
\Phi(x) = (Hx\cdot \xi)^\kappa .
\end{eqnarray}
Setting $\xi = (\xi^0, \boldsymbol\xi)$, with $\boldsymbol\xi = \|\boldsymbol\xi\|v\in{\mathbb{R}}^4$, $v\in S^3$, and $|\xi^0| = \|\boldsymbol\xi\|$, the dot product $Hx\cdot \xi$ takes the following form
\begin{eqnarray}
Hx\cdot \xi = (\tan \rho)\xi^0 - \frac{1}{\cos \rho}u\cdot \boldsymbol\xi = \frac{\xi^0e^{i\rho}}{2i\cos \rho} (1 + z^2 - 2zt),\nonumber
\end{eqnarray}
where $z =  ie^{-i\rho}\mbox{sgn}\;\xi^0$ and $t = u\cdot v \equiv \cos \varpi$. Now, utilizing the generating function for Gegenbauer polynomials,
\begin{eqnarray}\label{gefgpo}
(1 + z^2 - 2zt)^{-\lambda} = \sum_{n=0}^\infty z^n C_n^\lambda (t),
\end{eqnarray}
with $|z|<1$, and making use of this expression with $\lambda = -\kappa$, the following expansion can be established
\begin{eqnarray}\label{expdp}
(Hx\cdot \xi)^{\kappa}  =  \Big[ \Big( \frac{\xi^0e^{i\rho}}{2i\cos \rho} \Big)^{\kappa} \displaystyle\sum\limits_{n=0}^\infty z^nC_n^{-\kappa}(t) \Big], \; \Re\kappa <\frac{1}{2}.\hspace{.2cm}
\end{eqnarray}
Although (\ref{expdp}) is not valid in the sense of functions since $|z| = 1$, the convergence is ensured if we give a negative imaginary part to the angle $\rho$. In this regard, the ambient coordinates is extended to the forward tube \cite{BrosMoschella},
$$ {\cal{T}}^+ = \{ \mathbb{R}^5 - i\bar{V}_5^+ \cap M_H^{\mathbb{C}} \},\; \bar{V}_5^+ = \{ x\in \mathbb{R}^5: x^2\geq0,x^0>0 \}.$$

Now, we make use of two expansion formulas involving Gegenbauer polynomials \cite{Hua} and normalized hyperspherical harmonics on $S^3$, i.e.,
\begin{eqnarray}\label{gpo}
C_n^\lambda(t) = \frac{1}{\Gamma(\lambda)\Gamma(\lambda-1)}\displaystyle\sum\limits_{k=0}^{[\frac{n}{2}]}c_kC_{n-2k}^1(t),
\end{eqnarray}
with
\begin{eqnarray}
c_k = \frac{(n-2k+1)\Gamma(k+\lambda-1)\Gamma(\lambda+n-k)}{k!\Gamma(n-k+2)},\nonumber
\end{eqnarray}
and
\begin{eqnarray}\label{nohyha}
C_L^1(v\cdot v') = \frac{2\pi^2}{L+1}\displaystyle\sum\limits_{lm}Y_{Llm}(v)Y_{Llm}^*(v'),
\end{eqnarray}
in which $v,v'\in S^3$ and $(L,l,m)\in \mathbb{N}\times\mathbb{N}\times\mathbb{Z}$, with $0\leq l\leq L$ and $-l\leq m \leq l$. Combining the Gegenbauer polynomials generating function (\ref{gefgpo}) and two expansion formulas, (\ref{gpo}) and (\ref{nohyha}), we have
\begin{eqnarray}\label{expfo22}
(1+z^2-2zv\cdot v')^{-\lambda}\hspace{4cm} \nonumber\\
= 2\pi^2 \displaystyle\sum\limits_{Llm}z^Lp_L^\lambda(z^2)Y_{Llm}(v)Y_{Llm}^*(v'),
\end{eqnarray}
and the integral representation
\begin{eqnarray}\label{intrep1}
z^L p_L^\lambda(z^2)Y_{Llm}(v)\hspace{4.8cm} \nonumber\\
=\frac{1}{2\pi^2}\int_{S^3}(1+z^2-2zv\cdot v')^{-\lambda}Y_{Llm}(v^\prime)d\sigma (v^\prime),
\end{eqnarray}
where
\begin{eqnarray}
p_L^\lambda(z^2) = \frac{1}{(L+1)!}\frac{\Gamma(\lambda+L)}{\Gamma(\lambda)} \,_2F_1(L+\lambda,\lambda-1;L+2;z^2). \nonumber
\end{eqnarray}

Finally, let us return to the dS plane waves $(Hx\cdot\xi)^\kappa$. Considering (\ref{expdp}) and (\ref{expfo22}), the expansion of the dS plane waves becomes
\begin{eqnarray}\label{exdspwave}
(Hx\cdot\xi)^{\kappa}=2\pi^2\sum_{Llm}\Phi^{\kappa}_{Llm}(x)(\xi^0)^{\kappa}(\mbox{sgn}\;\xi^0)^LY_{Llm}^\ast(v),\;\;\;
\end{eqnarray}
where the functions
\begin{eqnarray}\label{SOF}
\Phi_{Llm}^{\kappa}(x) =\frac{i^{L-\kappa}e^{-i(L-\kappa)\rho}}{(2\cos \rho)^{\kappa}}p_L^{-\kappa}(-e^{-2i\rho})Y_{Llm}(u),\;\;\;
\end{eqnarray}
are introduced on the dS hyperboloid. By making use of the following relation between hypergeometric functions \cite{Magnus},
\begin{eqnarray}
_2F_1(a,b;c;z) = {(1-z)^{(c-a-b)}} {_2F_1}(c-a,c-b;c;z),\nonumber
\end{eqnarray}
the functions $\Phi_{Llm}^{\kappa}(x)$ take the form
\begin{eqnarray}
\Phi_{Llm}^{\kappa}(x) = i^{L-\kappa} e^{-i(L+\kappa +3)\rho} (2\cos \rho)^{\kappa +3}  \frac{\Gamma(L-\kappa)}{(L+1)! \Gamma(-\kappa)}\nonumber\\
\times {_2F_1}(\kappa +2,L+\kappa +3;L+2;-e^{-2i\rho})Y_{Llm}(u).\nonumber
\end{eqnarray}
Since $Y_{Llm}$'s are linearly independent, it is clear that $\Phi_{Llm}^{\kappa}(x)$'s are solutions to (\ref{Caopeq}) when adopting the appropriate separation of variables. In addition, the hyperspherical harmonics possess the important property that they are orthonormal. From it we get
\begin{eqnarray}\label{Fourier}
\Phi^{\kappa}_{Llm}(x) = \frac{(\mbox{sgn}\xi^0)^L}{2\pi^2 (\xi^0)^{\kappa}} \int_{S^3} d\sigma(v) (Hx\cdot\xi)^{\kappa}Y_{Llm}(v).\;\;\;\;
\end{eqnarray}
The above relations make explicit the `spherical' modes in dS spacetime in terms of the dS plane waves.

\setcounter{equation}{0}
\section{\label{bitensors}The two-point function from maximally symmetric bitensors in ambient space}
Following Allen and Jacobson in Ref. \cite{AllenJacobson}, any maximally symmetric bitensor can be presented as a sum of products of three basic tensors \cite{AllenJacobson}, while the expansion coefficients are determined by the geodesic distance $\sigma(x,x^\prime)$. [The geodesic distance is defined as the distance along the geodesic connecting the two points $x$ and $x^\prime$.] Of course, it can also be defined by a unique analytic extension when no geodesic connects $x$ and $x^\prime$. On this basis, the bitensors constitute a complete set. These fundamental tensors can be obtained by
\begin{eqnarray}
&n_\mu = \nabla_\mu \sigma(x, x^{\prime}),\;\;\; n_{\mu^{\prime}} = \nabla_{\mu^{\prime}} \sigma(x,x^{\prime}),&\nonumber\\
&g_{\mu\nu^{\prime}} = -c^{-1}({\cal{Z}})\nabla_{\mu}n_{\nu^{\prime}}+n_\mu n_{\nu^{\prime}}.&
\end{eqnarray}
The geodesic distance, for ${\cal{Z}}=-H^2x\cdot x^{\prime}$, can be implicitly given by \cite{BrosMoschella}
\begin{eqnarray}
\left\{\begin{array}{rl}
     {\cal{Z}}&=\cosh (H\sigma ), \;\hbox{if $x$ and $x^{\prime}$ are time-like separated,} \nonumber\\
     {\cal{Z}}&=\cos (H\sigma ), \;\hbox{if $x$ and $x^{\prime}$are space-like separated.} \nonumber\\
\end{array}\right.\;\;
\end{eqnarray}
The two-point function, with respect to these basis bitensors, then can be written as follows
\begin{eqnarray}
{\cal{W}}_{\mu\nu\mu^{\prime}\nu^{\prime}} = {\cal{A}}_1 (\sigma)g_{\mu\nu}g^{\prime}_{\mu^{\prime}\nu^{\prime}} + {\cal{A}}_2 (\sigma)g_{\mu\mu^{\prime}}g^{\prime}_{\nu\nu^{\prime}}\nn\\
+ {\cal{A}}_3 (\sigma)(g_{\mu\nu}n_{\mu^{\prime}}n_{\nu^{\prime}} + g^{\prime}_{\mu^{\prime}\nu^{\prime}}n_\mu n_\nu)\nn\\
+ {\cal{A}}_4 (\sigma)g_{\mu\mu^{\prime}}n_\nu n_{\nu^{\prime}} + {\cal{A}}_5 (\sigma)n_\mu n_\nu n_{\mu^{\prime}}n_{\nu^{\prime}}.
\end{eqnarray}
Since in the present paper we use the ambient space notations to evaluate the graviton two-point function, let us accordingly rewrite the basic bitensors based upon the associated notations.

In the ambient space formalism, we claim that the basic bitensors associated with $n_\mu$, $n_{\mu^\prime}$, and $g_{\mu\nu^\prime}$ can be, respectively, written as
\begin{eqnarray}
\bar{\partial}_\alpha \sigma(x,x^{\prime}),\;\;\;\bar{\partial}_{\beta^{\prime}}^{^{\prime}} \sigma(x,x^{\prime}),\;\;\;\theta_\alpha\cdot\theta^{\prime}_{\beta^{\prime}},
\end{eqnarray}
In order to prove this, one only should respect the restriction to the hyperboloid determined by
\begin{eqnarray}
{\cal{T}}_{\mu\nu^{\prime}}=x^{\alpha}_{\mu}{x^{\prime}}^{\beta^{\prime}}_{\nu^{\prime}}T_{\alpha\beta^{\prime}}.
\end{eqnarray}
\begin{itemize}
\item{When $ {\cal{Z}}=\cos(H\sigma)$, we have
\begin{eqnarray}
n_\mu &=& x^{\alpha}_{\mu}\bar{\partial}_\alpha \sigma(x,x^{\prime})= c({\cal{Z}}) x^{\alpha}_{\mu} (x^{\prime}\cdot\theta_\alpha),\nonumber\\
n_{\nu^{\prime}} &=& {x^{\prime}}^{\beta^{\prime}}_{\nu^{\prime}}\bar{\partial}_{{\beta^{\prime}}}^{^{\prime}} \sigma(x,x^{\prime}) = c({\cal{Z}}){x^{\prime}}^{\beta^{\prime}}_{\nu^{\prime}} (x\cdot\theta^{\prime}_{\beta^{\prime}}),\nonumber\\
\nabla_\mu n_{\nu^{\prime}} &=& x^{\alpha}_{\mu}{x^{\prime}}^{\beta^{\prime}}_{\nu^{\prime}}\theta^\varrho_\alpha{\theta^{\prime}}^{\gamma^{\prime}}_{\beta^{\prime}}\bar{\partial}_\varrho\bar{\partial}_{{\gamma^{\prime}}}^{^{\prime}} \sigma(x, x^{\prime}) \nonumber\\
&=& c({\cal{Z}})\Big[x^{\alpha}_{\mu}{x^{\prime}}^{\beta^{\prime}}_{\nu^{\prime}}\theta_\alpha \cdot\theta^{\prime}_{\beta^{\prime}}- n_\mu n_{\nu^{\prime}}{\cal{Z}}\Big],\nonumber
\end{eqnarray}
where $c({\cal{Z}})= {H}/{\sqrt{1-{\cal{Z}}^2}}$, ${x}^{\alpha}_{\mu} = {\partial{x}^{\alpha}}/{\partial {X^{\mu}}}$ and ${x^{\prime}}^{\beta^{\prime}}_{\nu^{\prime}} = {\partial {x^{\prime}}^{\beta^{\prime}}}/{\partial {X^{\prime}}^{\nu^{\prime}}}$.}
\item{When ${\cal{Z}}=\cosh (H\sigma)$, $n_\mu$ and $n_{\nu^{\prime}}$ are multiplied by $i$ and $c({\cal{Z}})={iH}/{\sqrt{1-{\cal{Z}}^2}}$. For both cases, one finds
\begin{eqnarray}
g_{\mu\nu^{\prime}}+({\cal{Z}}-1)n_\mu n_{\nu^{\prime}}=x^{\alpha}_{\mu}{x^{\prime}}^{\beta^{\prime}}_{\nu^{\prime}}\theta_\alpha \cdot\theta^{\prime}_{\beta^{\prime}}.\nonumber
\end{eqnarray}}
\end{itemize}

\setcounter{equation}{0}
\section{\label{Some useful relations}Some useful relations}
To obtain the two-point function, the following identities are used
\begin{eqnarray}
\partial_2\cdot D_2 {D^{\prime}_2}{\cal{W}}_g = -(Q_1 + 6) {D^{\prime}_2}{\cal{W}}_g,
\end{eqnarray}
\begin{eqnarray}
Q_2D_2 {D^{\prime}_2}{\cal{W}}_g = D_2 Q_1 {D^{\prime}_2}{\cal{W}}_g,
\end{eqnarray}
\begin{eqnarray}
Q_2 \theta\theta^{\prime}{\cal{W}}_0 = \theta Q_0 \theta^{\prime}{\cal{W}}_0,
\end{eqnarray}
\begin{eqnarray}
\partial_2\cdot\theta \theta^{\prime}{\cal{W}}_0 = -H^{2} D_1 \theta^{\prime}{\cal{W}}_0,
\end{eqnarray}
\begin{eqnarray}
&Q_2{\cal{S}}{\cal{S}}^\prime \theta\cdot\theta^{\prime}{\cal{W}}_1= {\cal{S}}{\cal{S}}^\prime\theta\cdot\theta^{\prime}(Q_1-4){\cal{W}}_1 \hspace{2cm}&\nonumber\\
&\hspace{1cm}-2H^2{\cal{S}}^\prime D_2x\cdot\theta^\prime{\cal{W}}_1 + 4{\cal{S}}^\prime\theta\theta^\prime\cdot{\cal{W}}_1,
\end{eqnarray}
\begin{eqnarray}
&\partial_2\cdot {\cal{S}}{\cal{S}}^\prime \theta\cdot\theta^{\prime}{\cal{W}}_1 =
T {\cal{S}}^\prime\theta^\prime\cdot\bar\partial{\cal{W}}_1\hspace{2cm}&\nonumber\\
&\hspace{1cm} - H^{2}{\cal{S}}^\prime D_1 \theta^\prime\cdot{\cal{W}}_1 + 5 H^{2}{\cal{S}}^\prime x\cdot\theta^\prime {\cal{W}}_1,&
\end{eqnarray}
\begin{eqnarray}
\bar{\partial}_\alpha f({\cal{Z}})=-(x^{\prime}\cdot\theta_{\alpha})\frac{d f(\cal{Z})}{d{\cal{Z}}},
\end{eqnarray}
\begin{eqnarray}
\theta^{\alpha\beta}\theta^{\prime}_{\alpha\beta}=\theta\cdot\cdot\theta^{\prime}=3+{\cal{Z}}^2,
\end{eqnarray}
\begin{eqnarray}
(x.\theta^{\prime}_{\alpha^{\prime}})(x\cdot{\theta^{\prime}}^{\alpha^{\prime}})={\cal{Z}}^2-1,
\end{eqnarray}
\begin{eqnarray}
(x.\theta^{\prime}_{\alpha})(x^{\prime}\cdot\theta^{\alpha})={\cal{Z}}(1-{\cal{Z}}^2),
\end{eqnarray}
\begin{eqnarray}
\bar{\partial}_\alpha(x\cdot\theta^{\prime}_{\beta^{\prime}})=\theta_{\alpha}\cdot\theta^{\prime}_{\beta^{\prime}},
\end{eqnarray}
\begin{eqnarray}
 \bar{\partial}_\alpha(x^{\prime}\cdot\theta_{\beta})=x_\beta(x^{\prime}\cdot\theta_{\alpha})-{\cal{Z}}\theta_{\alpha\beta},
\end{eqnarray}
\begin{eqnarray}
\bar{\partial}_\alpha(\theta_{\beta}\cdot\theta^{\prime}_{\beta^{\prime}})=x_\beta(\theta_{\alpha}\cdot\theta^{\prime}_{\beta^{\prime}})+ \theta_{\alpha\beta}(x\cdot\theta^{\prime}_{\beta^{\prime}}),
\end{eqnarray}
\begin{eqnarray}
{\theta^{\prime}}^{\beta}_{\alpha^{\prime}}(x^{\prime}\cdot\theta_{\beta})=-{\cal{Z}}(x\cdot\theta^{\prime}_{\alpha^{\prime}}),
\end{eqnarray}
\begin{eqnarray}
{\theta^{\prime}}^{\gamma}_{\alpha^{\prime}}(\theta_{\gamma}\cdot\theta^{\prime}_{\beta^{\prime}})=\theta^{\prime}_{\alpha^{\prime}\beta^{\prime}}+(x\cdot\theta^{\prime}_{\alpha^{\prime}})(x\cdot\theta^{\prime}_{\beta^{\prime}}),
\end{eqnarray}
\begin{eqnarray}\label{A.16}
Q_0f({\cal{Z}})=(1-{\cal{Z}}^2)\frac{d^2 f(\cal{Z})}{d{\cal{Z}}^2}-4{\cal{Z}}\frac{d f(\cal{Z})}{d{\cal{Z}}}.
\end{eqnarray}

\end{appendix}

\end{document}